\documentclass[12pt]{article}

\usepackage{sbc-template}

\usepackage{graphicx,url}

\usepackage[latin1]{inputenc}  
\usepackage{float}
\usepackage{amsmath}
\usepackage{amssymb}
\usepackage{amsfonts}
\usepackage{euscript}
\usepackage{graphicx}
\usepackage[T1]{fontenc}
\usepackage[portuges,english]{babel}

\newcounter{lemma}
\newcounter{corollary}
\newcounter{proposition}
\newcounter{definition}
\newcounter{example}
\newcounter{remark}
\newcounter{question}
\newcounter{notation}

\newtheorem{lemma}{Lemma}          

\newtheorem{definition}{Definition}

\title{A SCHEME OF CONCATENATED QUANTUM CODE TO PROTECT AGAINST BOTH COMPUTATIONAL ERROR AND AN ERASURE}

\author{G. O. Santos$^{1,3}$, F. M.  Assis$^{1,3}$, A. F. Lima$^{2,3}$}

\address{Department of Electrical Engineering\\
$^{2}$Department of Physics\\
$^{3}$Institute for Studies on Quantum Computation and Quantum Information (IQuanta)\\
Federal University of Campina Grande\\
Rua Aprígio Veloso, 882 -- Campina Grande -- Paraíba -- Brazil\\
gilson.santos@ee.ufcg.edu.br, fmarcos@dee.ufcg.edu.br, aerlima@df.ufcg.edu.br}

\DeclareMathAlphabet{\mathpzc}{OT1}{pzc}{m}{it}
\begin{document} 

\maketitle

\begin{abstract}
We present a description of encoding/decoding for a concatenated quantum code that enables both protection against quantum computational errors and the occurrence of one quantum erasure. For this, it is presented how encoding and decoding for quantum graph codes are done, which will provide the protection against the occurrence of computational errors (external code). As internal code is used encoding and decoding via scheme of GHZ states for protection against the occurrence of one quantum erasure.
\end{abstract}


\section{Introduction}	
Schlingemann and Werner \cite{1} have presented a new way to construct quantum stabilizer codes using graphs with specific properties, called quantum graph codes. They also established the necessary and sufficient conditions for a graph to generate a quantum error-correcting code (QECC), adapted of Knill and Laflamme\cite{2}. Feng\cite{3} presents in other way this conditions established from Schlingemann and Werner for a graph to generate a QECC. 

Grassl et al.\cite{4} considered a situation in which the position of the erroneous qubits is known. According to classical coding theory, they called this model the quantum erasure channel (QEC).

\indent Alterations or changes caused by the environment can be characterized as being of two types: (i) those that satisfy to certain conditions that allow their correction, i.e., that agree with the conditions established by Knill and Laflamme\cite{2}. They are represented by Pauli matrices and are called ``computational errors''\cite{5}. Such matrices constitute the computational space, and (ii) those that lead the state encoded out of the computational space. These alterations characterize the QEC.

\indent Erasures are both detectable and locatable, which suggests that they should be ea\-sier to rectify than computational errors. In fact, quantum communication channels can to\-le\-ra\-te a higher rate of erasure ($p_{erasure} < 0.5$) than depolarization ($p_{comp} <1/3$)\cite{6}. Dawson et al.\cite{7} considered an error model which contains both erasure and computational errors, finding that fault-tolerant quantum computation is possible with $p_{erasure} < 3 \times 10^{-3}$ and $p_{comp} < 10^{-4}$.

\indent In this paper we present a construction of a quantum code that ena\-bles protection against both types of changes caused by the environment. For this, we cosidered as internal code for protection against the occurrence of one erasure the encoding and decoding via Greenberger-Horne-Zeilinger (GHZ) states, for this operators have been constructed by generalizing the operators given by Yang et al\cite{8}. As extenal code,  we presents how encoding and decoding for quantum graph codes are done, for this we developed a decoding operator making use the inverse quantum Fourier transform. This code (external code) will provide the protection against the occurrence of computational errors. 

\indent In Section \ref{prelim} includes some introductory remarks about alterations caused by the environment and some comments on the model of quantum erasure channel. In Section \ref{cgraphquant} there is a description of the encoding/decoding for quantum graph codes. In Section \ref{erasurecode} is presented a generalization of scheme of erasure-correcting code via GHZ states. In Section \ref{conscodigo} it is shown  the realization of concatenated code making use of a quantum graph code and a scheme via GHZ state. Finally, in Section \ref{conclus} final comments regarding this work are made.\\

\section{Preliminaries}\label{prelim}
It is usually assumed that the space system $\mathcal{H}_{sys}$ is a tensor product of two-dimensional spaces $\mathcal{H}_{2}$ (qubits), i.e.,
{\footnotesize
$$
\mathcal{H}_{sys} = \mathcal{H}_{2} \otimes \ldots \otimes \mathcal{H}_{2} . 
$$
}

However, this is an approximation\cite{4}. For example, atoms usually have many le\-vels that can be occupied, it is assigned to an unwanted dynamical evolution of the system. Therefore, the Hilbert space of the system $\mathcal{H}_{sys}$ is a tensor product of multidimensional spaces with two-dimensional subspaces used for computing:
{\footnotesize
$$
\mathcal{H}_{sys} = \mathcal{H}_{k} \otimes \ldots \otimes \mathcal{H}_{k} ,
$$
$$
\mathcal{H}_{comp} = \mathcal{H}_{2} \otimes \ldots \otimes \mathcal{H}_{2} ,
$$
}
where $\mathcal{H}_{comp}$ is the subspace of allowed computational states. Each two-dimensional space $\mathcal{H}_{2}$ is a subspace of $\mathcal{H}_{k}$, but not necessarily a tensor factor of $\mathcal{H}_{k}$ (for simplicity we assume that the dimensions of all tensor factors are equal). Therefore, the system space $\mathcal{H}_{sys}$ can only be decomposed as a direct sum of subspaces
{\footnotesize
$$
\mathcal{H}_{sys} = \mathcal{H}_{comp} \oplus \mathcal{H}^{\perp}_{comp}
$$
}
and generally not as a tensor product. During the free-erasure computations the system remains in $\mathcal{H}_{comp}$. Any population found in $\mathcal{H}^{\perp}_{comp}$ is the signature of erasure \cite{4}. 

\indent The QEC model is illustrated in the following situation, see Ref.~ \cite{9}. Consider a physical system consisting of an atom in which the information is properly encoded into two energy levels. These levels constitute the computational space $\mathcal{H}_{comp}$. A transition that takes a system state out of the computational space $\mathcal{H}^{\perp}_{comp}$ is then called a non-resonant transition. The manipulation of the system  is done through techniques, using a laser tuned in the difference of energy between these two states (resonant transition), which allow to identify the population of the levels of work. If a transition is signaled by the emission of a photon, for example, and the measurement of the population of the predicted level does not indicate variation, it is concluded that there was a non-resonant transition. The state was thus taken out of the computational space. This fact characterizes the quantum erasure channel.

\indent Furthermore, it is believed that, although stimulated transitions are more frequent than spontaneous in the microwave band (frequency $\nu \sim 2 \times 10^{10} Hz$), but spontaneous transitions are more frequent than sti\-mu\-la\-ted in the optical domain (frequency $\nu \sim 6 \times 10^{14} Hz$) \cite{10}. It should be noted that the optics are currently the domain for quantum communication, which makes the occurrence of erasure quite relevant for this type of communication.

\indent If there is the possibility of erasure detection, then it is not necessary to use a code that protects against computational errors to perform the correction of error in the erasure channel. To do this, we can use a code that performs only protection against erasure \cite{1,11}. Seeing this and what was stated above, it is considered here a concatenated code in which the internal encoding and internal decoding are built to protect against erasure.\\
\indent In general, alteration of information is not {\it a priori} obvious for the observer, which should encode the information in a special way to detect such change. One way that can be explored to perform this encoding is the use of Greenberger-Horne-Zeilinger (GHZ) state \cite{12,13}. The GHZ state is a maximally entangled state of three qubits. But, despite the fact that the GHZ state has been idealized for three qubits, it can be extended for N-qubits, see Refs. \cite{14}-\cite{16}. Because of this, the GHZ state has been useful for many applications, including the construction of QECC for multidimensional systems.

\section{Decoding for Quantum Graph Codes}\label{cgraphquant}
In this section we present a way to accomplish the decoding for quantum graph codes. This code will protect the encoded information against the occurrence of quantum computational errors.

\indent The description of a graph (or matrix) for quantum codes, given by Schlingemann and Werner \cite{1}, can be stated as follows.
\begin{definition}\label{ingred}
{\it Let $X$ and $Y$ be disjoint sets with $k$ and $n$ elements, respectively, where the set $X$ represents the input vertices and the set $Y$ represents the output vertices. Consider $G = (V(G), E(G))$ a graph with a set of vertices $V = V (G) = X \cup Y$ and a set of edges $E = E(G) \subset V \times V$. Each vertice represents a qubit, and each edge $\overline{v_{i}v_{j}} \in E (v_{i}, v_{j} \in V)$, assigned a weight $a_{v_{i}v_{j}} \in \mathbb{F}_{p}$, represents the interaction between the qubits. The graph has to be undirected and simple. The matrix $A = ( a_{v_{i} v_{j}} )$ of a graph $G = (V, E)$ corresponds to a simetric array (adjacency matrix) over $\mathbb{F}_{p}$ with dimensions $(n + k) \times (n + k)$, where $\mathbb{F}_{p}$ is a finite Abelian group with $p$ (prime) elements. }
\end{definition}

\indent We fix a prime number $p$. A vector in the vector space $\mathbb{F}_{p}^{|V|} = \mathbb{F}_{p}^{n + k}$ is represented by a column vector 
{\footnotesize
$$
d^{V} = \left( \begin{array}{c} d^{v_{1}} \\ \vdots \\ d^{v_{n+k}} \end{array} \right) = \{ d^{v} | v \in V \}
$$
}
where $V=\{ v_{1}, \ldots, v_{n + k} \}$ and $d^{v_{i}} \in \mathbb{F}_{p}$. For a subset $S$ of $V$ will be denoted 

{\footnotesize
$$
d^{S} =  \{ d^{s} | s \in S \} \in \mathbb{F}_{p}^{|S|},
$$
}
\\
the column vector equivalent. 

\indent A qubit $\vert v \rangle$ is a nonzero vector in $\mathbb{C}^{p}$
{\footnotesize
\begin{equation}
\vert v \rangle = c_{0} \vert 0 \rangle + c_{1} \vert 1 \rangle + \ldots +c_{p-1} \vert p - 1 \rangle \ \ \ (c_{i} \in \mathbb{C} )
\end{equation}
}
where 
{\footnotesize
\begin{equation}\label{vbasis}
\{ \vert 0 \rangle , \vert 1 \rangle , \ldots , \vert p - 1 \rangle \} = \{ \vert a \rangle : a \in \mathbb{F}_{p} \}
\end{equation}
}
is a basis of $\mathbb{C}^{p}$. An $n$-qubit is a nonzero vector in $(\mathbb{C}^{p})^{\otimes n}=(\mathbb{C}^{p})^{n}$. We choose  a basis of $(\mathbb{C}^{p})^{n}$ by 
{\footnotesize
\begin{equation}\label{defnqubit}
\{ \vert a_{1} \ldots a_{n} \rangle = \vert a_{1} \rangle \otimes \vert a_{2} \rangle \otimes \ldots \otimes \vert a_{n} \rangle : (a_{1}, \ldots , a_{n} ) \in \mathbb{F}_{p}^{n} \}
\end{equation}
}
so that an $n$-qubit can be expressed as a nonzero vector
{\footnotesize
\begin{equation}
\vert v \rangle = \sum_{a = (a_{1}, \ldots , a_{n}) \in \mathbb{F}_{p}^{n}} c_{a} \vert a \rangle = \sum_{(a_{1}, \ldots , a_{n}) \in \mathbb{F}_{p}^{n}} c_{a_{1} \ldots a_{n} } \vert a_{1} \ldots a_{n} \rangle \ \ \ (c_{a} = c_{a_{1} \ldots a_{n} } \in \mathbb{C} ) .
\end{equation}
}
\indent We present the following as the quantum error-correcting codes associated with graphs (called {\it quantum graph codes}) are encoding.

\begin{lemma}\label{defcodgraf}\cite{3}
{\it Consider a graph satisfying the description given in Definition \ref{ingred}. We define the $\mathbb{C}$-linear mapping \\
{\footnotesize
\begin{eqnarray}
f : \ \ \ (\mathbb{C}^{p})^{\otimes k} & \rightarrow & (\mathbb{C}^{p})^{\otimes n}  \nonumber \\
     \vert v \rangle                      & \mapsto    & f( \vert v \rangle ) 
\end{eqnarray}
}
where for an element
{\footnotesize
\begin{eqnarray}
\vert v \rangle = \sum_{d^{X} \in \mathbb{F}_{p}^{|X|}} c(d^{X}) \vert d^{X} \rangle & \in (\mathbb{C}^{p})^{\otimes k} \\
                                                                      & (c(d^{X}) \in \mathbb{C} ) \nonumber
\end{eqnarray}
}
we define
{\footnotesize
\begin{equation}\label{codifgrafo}
f( \vert v \rangle ) = \frac{1}{\sqrt{p^{|Y|}}} \sum_{d^{Y} \in \mathbb{F}_{p}^{|Y|}} \lambda (d^{Y}) \vert d^{Y} \rangle \ \in (\mathbb{C}^{p})^{\otimes n} .
\end{equation}
}
where 
{\footnotesize
\begin{equation}\label{lambdadef}
\lambda (d^{Y}) = \sum_{d^{X} \in \mathbb{F}_{p}^{|X|}} e^{(\frac{2\pi i}{p}) \left( \frac{1}{2}((d^{X})^{t},(d^{Y})^{t}).A.\left({\tiny \begin{array}{c} d^{X} \\ d^{Y} \end{array}} \right) \right)} . c(d^{X}) \in \mathbb{C}.
\end{equation}
}
Let $Q$ be the image of $f$, the set $Q=\{ f(\vert v \rangle ) : \ \vert v \rangle \in (\mathbb{C}^{p})^{\otimes k} \}$ is a quantum code $[[ n, k, d ]]_{p}$.
}
\end{lemma}

\indent The proof of Lemma \ref{defcodgraf} is given in the Appendix of Ref. \cite{3}. 

\indent The matrix A of Eq. (\ref{lambdadef}) have the following subdivision
{\footnotesize
\begin{equation}\label{mtxadjc1}
A = \left( 
\begin{array}{c  c} 
A_{XX} & A_{XY} \\
A_{YX} & A_{YY} \\
\end{array}
\right) .
\end{equation}
}
\indent When the encoding is done, we obtain
{\footnotesize
\begin{equation}
\vert \psi \rangle =  \frac{1}{\sqrt{p^{|Y|}}} \left[ \lambda (d^{Y(1)}) . \vert d^{Y(1)} \rangle + \lambda (d^{Y(2)}) . \vert d^{Y(2)} \rangle + \ldots + \lambda (d^{Y(p^{|Y|})}) . \vert d^{Y(p^{|Y|})} \rangle \right]
\end{equation}
}
or compactly represented by
{\footnotesize
\begin{equation}\label{encodgph}
\vert \psi \rangle = \frac{1}{\sqrt{p^{|Y|}}} \sum_{j=1}^{p^{|Y|}} \lambda (d^{Y(j)}) \vert d^{Y(j)} \rangle
\end{equation}
}
where the integer $j$ corresponding to its decomposition in $p$. To simplify the notation, from this point on we will omit the normalization factors.

\indent A quantum computational error $\sigma_{b}\tau_{s} \ (b, s \in \mathbb{F}_{p})$ on a qubit is a unitary linear operator on $\mathbb{C}^{p}$, which acts on the basis (\ref{vbasis}) of $\mathbb{C}^{p}$ by
{\footnotesize
\begin{equation}
\sigma_{b}\tau_{s} \vert a \rangle = e^{\frac{2 \pi i}{p} (s a) } \vert a + b \rangle \ \ \  (a \in \mathbb{F}_{p} ).
\end{equation}
}
The set 
{\footnotesize
\begin{equation}
\mathpzc{E}_{1} = \{ e^{(\frac{2 \pi i}{p}) m}\sigma_{b} \tau_{s} | m,b,s \in \mathbb{F}_{p} \}
\end{equation}
}
forms a (non-Abelian) error group \cite{3}. A quantum computational error on an $n$-qubit is a unitary linear operator on $(\mathbb{C}^{p})^{\otimes n}$
{\footnotesize
\begin{equation}\label{errodef}
\varepsilon = e^{(\frac{2 \pi i}{p}) m} \omega_{1} \otimes \omega_{2} \otimes \ldots \otimes \omega_{n} \ \ \ \ (\omega_{j} = \sigma_{j} \tau_{j}; m, b_{j}, s_{j} \in \mathbb{F}_{p} )
\end{equation}
}
which acts on the basis (\ref{defnqubit}) of $(\mathbb{C}^{p})^{n}$ by
{\footnotesize
\begin{eqnarray}\label{nerror}
\varepsilon \vert a_{1} \ldots a_{n} \rangle & = & e^{(\frac{2 \pi i}{p}) m} . ( \omega_{1}  \vert a_{1} \rangle ) \otimes (\omega_{2} \vert a_{2} \rangle ) \otimes \ldots \otimes (\omega_{n} \vert a_{n} \rangle ) \nonumber \\ 
 & = &  e^{(\frac{2 \pi i}{p}) m} . e^{(\frac{2 \pi i}{p}) (s_{1}a_{1} + s_{2}a_{2} + \ldots + s_{n} a_{n})} (\vert a_{1} + b_{1} \rangle \otimes \vert a_{2} + b_{2} \rangle \otimes \ldots \otimes \vert a_{n} + b_{n} \rangle ) \nonumber \\
& = &  e^{(\frac{2 \pi i}{p}) m} .  e^{(\frac{2 \pi i}{p}) [(s)^{t}.(a)]} \vert a + b \rangle \ \ \ [s = (s_{1}, \ldots , s_{n}), b = (b_{1}, \ldots , b_{n} ) \in \mathbb{F}_{p}^{n} ].  
\end{eqnarray} 
}
where $(s)^{t}$ represents the transpose of the vector $(s)$. The set of all errors $\varepsilon$ of (\ref{errodef}) forms an error group $\mathpzc{E}_{n}$. 

\indent For such $\varepsilon$ of the form (\ref{errodef}), we can find $\mathpzc{E} \subseteq Y$, $I=Y \setminus \mathpzc{E}$ such that
{\footnotesize
\begin{equation}
\varepsilon \vert d^{Y} \rangle = \varepsilon \vert d^{\mathpzc{E}} , d^{I} \rangle = e^{(\frac{2 \pi i}{p} ) [ (s^{\mathpzc{E}})^{t}. (d^{\mathpzc{E}}) ]} \vert d^{\mathpzc{E}} + b^{\mathpzc{E}}, d^{I} \rangle \ \ \ (s^{\mathpzc{E}}, b^{\mathpzc{E}} \in \mathbb{F}_{p}^{\mathpzc{E}}).
\end{equation}
}
Therefore, when the $\vert \psi \rangle$ state in (\ref{encodgph}) suffer an error specified by Eq. (\ref{errodef}) is produced 
{\footnotesize
\begin{eqnarray}\label{errorstate}
\vert \varphi \rangle = \varepsilon \vert \psi \rangle & = & \sum_{j=1}^{p^{|Y|}} \lambda (d^{Y(j)}) \varepsilon \vert d^{Y(j)} \rangle \nonumber \\
 & = & \sum_{j=1}^{p^{|Y|}} \lambda (d^{\mathpzc{E}(j)} , d^{I(j)}) . \varepsilon . \vert d^{\mathpzc{E}(j)}, d^{I(j)} \rangle \nonumber \\
 & = &  \sum_{j=1}^{p^{|Y|}} \lambda (d^{\mathpzc{E}(j)} , d^{I(j)}) . e^{(\frac{2 \pi i}{p}) [(s^{\mathpzc{E}(j)})^{t}.(d^{\mathpzc{E}(j)})]} . \vert d^{\mathpzc{E}(j)} + b^{\mathpzc{E}(j)}, d^{I(j)} \rangle \nonumber \\
 & = &  \sum_{j=1}^{p^{|Y|}} \lambda (d^{\mathpzc{E}(j)} - b^{\mathpzc{E}(j)}, d^{I(j)}) .  e^{(\frac{2 \pi i}{p}) [(s^{\mathpzc{E}(j)})^{t}.(d^{\mathpzc{E}(j)})]} . \vert d^{\mathpzc{E}(j)}, d^{I(j)} \rangle 
\end{eqnarray}
}
where 
{\footnotesize
\begin{equation}\label{errorstate2}
\lambda (d^{\mathpzc{E}(j)} - b^{\mathpzc{E}(j)}, d^{I(j)}) = \sum_{d^{X} \in \mathbb{F}_{p}^{|X|}} e^{(\frac{2 \pi i}{p})[\eta (j)]} . c(d^{X}) 
\end{equation}
}
and
{\footnotesize
\begin{equation}\label{errorstate3}
\eta (j) = \left\{ \frac{1}{2}[(d^{X})^{t},(d^{\mathpzc{E}(j)} - b^{\mathpzc{E}(j)})^{t},(d^{I(j)})^{t}].A.\left[{\tiny \begin{array}{c} d^{X} \\ d^{\mathpzc{E}(j)} - b^{\mathpzc{E}(j)} \\ d^{I(j)} \end{array}} \right] \right\} .
\end{equation}
}

\indent The matrix $A$ of Eq. (\ref{errorstate3}) has the following form
{\footnotesize
\begin{equation}\label{mtxadjc2}
A = \left( 
\begin{array}{c  c  c} 
A_{XX} & A_{X\mathpzc{E}} & A_{XI} \\
A_{\mathpzc{E}X} & A_{\mathpzc{EE}} & A _{\mathpzc{E}I} \\
A_{IX}  &  A_{I\mathpzc{E}}  & A_{II} \\
\end{array}
\right).
\end{equation}
}
Replacing the matrix form (\ref{mtxadjc2}) in the Eq. (\ref{errorstate3}) and considering that the submatrices $A_{XX}=0$ (by Def. \ref{ingred}), $A_{X\mathpzc{E}}=(A_{\mathpzc{E}X})^{t}$, $A_{XI}=(A_{IX})^{t}$ and $A_{\mathpzc{E}I}=(A_{I\mathpzc{E}})^{t}$.  After performing the calculations, we obtain
{\footnotesize
\begin{eqnarray}\label{errorstate4}
\eta (j) & = & (d^{X})^{t}A_{X\mathpzc{E}}(d^{\mathpzc{E}(j)}) - (d^{X})^{t}A_{X\mathpzc{E}}(b^{\mathpzc{E}(j)}) + (d^{X})^{t}A_{XI}(d^{I(j)}) + (d^{\mathpzc{E}(j)})^{t}A_{\mathpzc{E}I}(d^{I(j)}) \nonumber \\
         &   & - (b^{\mathpzc{E}(j)})^{t}A_{\mathpzc{E}I}(d^{I(j)}) + \frac{1}{2} (d^{\mathpzc{E}(j)})^{t}A_{\mathpzc{EE}}(d^{\mathpzc{E}(j)}) - \frac{1}{2}(d^{\mathpzc{E}(j)})^{t}A_{\mathpzc{EE}}(b^{\mathpzc{E}(j)}) - \frac{1}{2}(b^{\mathpzc{E}(j)})^{t}A_{\mathpzc{EE}}(d^{\mathpzc{E}(j)}) \nonumber \\
         &  &  + \frac{1}{2}(b^{\mathpzc{E}(j)})^{t}A_{\mathpzc{EE}}(b^{\mathpzc{E}(j)}) + \frac{1}{2}(d^{I(j)})^{t}A_{II}(d^{I(j)}) .
\end{eqnarray}
}

\indent It is important to note that in Eq. (\ref{errorstate}) the element responsible for bit-flip $b^{\mathpzc{E}(j)}$ is now represented in the exponential form in $\lambda (d^{\mathpzc{E}(j)} - b^{\mathpzc{E}(j)}, d^{I(j)})$ - Eqs. (\ref{errorstate2}) and (\ref{errorstate3}). Thus both the elements of phase-flip and bit-flip are in the exponential form.

\indent The decoding operation of a quantum graph code is based on an appropriate extension of the encoding graph by adding syndrome vertices $L$ and edges connecting these vertices with the output vertices $Y$ in an appropriate way.\cite{17} Syndrome vertices $L$ are vertices of measure used to establish the syndrome. The set of $e$-error correcting graphs with input vertices X, output vertices Y an syndrome vertices L is denoted by $\mathcal{G}_{e}(X,Y,L)$. Insertion of these syndrome vertices $L$ must satisfy the admissibility conditions established by Schlingemann \cite{17,18}. 

\begin{definition}\label{condgraphdecod}
{\it The set of graphs of correcting $e$-errors $\mathcal{G}_{e}(X,Y,L)$ is defined as consisting of all graphs on the union of the input vertices $X$, output vertices $Y$ and syndrome vertices $L$ for which the adjacency matrix $A = (A_{x}^{y})_{x,y \in XYL}$ satisfies the following conditions: 
\begin{description}
\item[(C1)] The sets of vertices satisfying the expression $|X|+|L|=|Y|$, where $| . |$ indicates the cardinality of the set;
\item[(C2)] The matrix $\overline{A}=A_{XY}^{L}$ is invertible;
\item[(C3)] There are no edges that connect syndrome vertices, i.e., $A_{LL}=0$;
\item[(C4)] There are no edges that connect input vertices and syndrome vertices, i.e., $A_{XL}=0$ and $A_{LX}=0$;
\item[(C5)] For all sets $\mathpzc{E} \subset Y$ containing at least a maximum of $2e$ graph elements, the graph must satisfy the necessary and sufficient conditions for a graph to generate an $e$-error correcting code, i.e., for $ I=Y \setminus \mathpzc{E}$
\end{description}
{\footnotesize
\begin{equation}\label{cond5}
 \mbox{if  } \ d^{X} \in \mathbb{F}_{p}^{|X|} \ \mbox{ and } d^{\mathpzc{E}} \in \mathbb{F}_{p}^{|\mathpzc{E}|} \mbox{  and } \  A_{IX} d^{X} + A_{I \mathpzc{E}} d^{\mathpzc{E}} = 0^{I} \ \mbox{implies} \ d^{X}=0^{X} \ \mbox{and} \ A_{X \mathpzc{E}} d^{\mathpzc{E}} = 0^{X} .
\end{equation}
}
}
\end{definition}

\indent The syndrome vertices $L$ are to be  seen as qudits which are measured in order  to fix the {\it error syndrome}. The error syndrome is the result of the measurement which tells us which error has occurred and which correction we have to apply. 

\indent Besides the graph satisfying the Definition \ref{condgraphdecod} and observing the Eqs. (\ref{errorstate}) e (\ref{errorstate2}), we developing the decoding ope\-ra\-tion for quantum graph codes which will be given by the following Lemma. 

\begin{lemma}\label{graphdecod}
{\it Making use of a graph $\mathcal{G}_{e}(X,Y,L)$ that attends to Definition \ref{condgraphdecod}, the decoding operation for the $\vert \varphi \rangle$ state is performed by 
{\footnotesize
\begin{equation}\label{decodgraph}
\mathbf{\mathcal{R}} (\vert \varphi \rangle ) = \frac{1}{\sqrt{p^{|\mathpzc{E}|+|I|}}} \sum_{j=1}^{p^{|Y|}} \lambda (d^{\mathpzc{E}(j)} - b^{\mathpzc{E}(j)}, d^{I(j)}) . \left( e^{(\frac{2 \pi i}{p}) [(s^{\mathpzc{E}(j)})^{t}.(d^{\mathpzc{E}(j)})]} \right) \mathbf{\mathcal{T}} ( \vert d^{\mathpzc{E}(j)}, d^{I(j)} \rangle ) 
\end{equation}
}
where
{\footnotesize
\begin{equation}\label{iqftgrafo}
\mathbf{\mathcal{T}} ( \vert d^{E(j)}, d^{I(j)} \rangle ) = \frac{1}{\sqrt{p^{|\mathpzc{E}|+|I|}}} \sum_{d^{L} \in \mathbb{F}_{p}^{L}}  \sum_{d^{X} \in \mathbb{F}_{p}^{X}}  e^{-(\frac{2\pi i}{p}) \left[ \mu (j) \right] } \vert d^{L}d^{X} \rangle,
\end{equation}
}
and 
{\footnotesize
\begin{equation}\label{expmu}
\mu (j) = \frac{1}{2} \left[  (d^{X})^{t}, (d^{\mathpzc{E}(j)})^{t}, (d^{I(j)})^{t}, (d^{L})^{t} \right].  \overline{A} . \left[ \begin{array}{c} d^{X} \\ d^{\mathpzc{E}(j)} \\ d^{I(j)} \\ d^{L} \end{array} \right]
\end{equation}
}
and $\lambda (d^{\mathpzc{E}(j)} - b^{\mathpzc{E}(j)}, d^{I(j)})$ is given by the Eq. (\ref{errorstate2}).
}
\end{lemma}
{\bf Proof --}
We will show that the operation defined by  $\mathcal{R}$ (Eq. (\ref{decodgraph})) identify the occurrence of a computational error as described in Eq. (\ref{errodef}) by combining inverse quantum Fourier transform (IQFT) $\mathcal{T}$  with the function $\lambda$ in Eq. (\ref{errorstate2})  for all basis states on the $\vert \varphi \rangle$  state. To simplify the notation, we will omit the normalization factors. 

\indent Let $\lambda (d^{\mathpzc{E}(j)} - b^{\mathpzc{E}(j)} , d^{I(j)})$ be of the Eq.(\ref{errorstate2}), i.e.,  
{\footnotesize
\begin{equation}\label{errorstate2b}
\lambda (d^{\mathpzc{E}(j)} - b^{\mathpzc{E}(j)}, d^{I(j)}) = \sum_{d^{X} \in \mathbb{F}_{p}^{|X|}} e^{(\frac{2 \pi i}{p}).[\eta (j)] } . c(d^{X}) 
\end{equation}
}
where
{\footnotesize
\begin{eqnarray}\label{eta3}
\eta (j) & = & (d^{X})^{t}A_{X\mathpzc{E}}(d^{\mathpzc{E}(j)}) - (d^{X})^{t}A_{X\mathpzc{E}}(b^{\mathpzc{E}(j)}) + (d^{X})^{t}A_{XI}(d^{I(j)})  \nonumber \\
         &   & + (d^{\mathpzc{E}(j)})^{t}A_{\mathpzc{E}I}(d^{I(j)}) - (b^{\mathpzc{E}(j)})^{t}A_{\mathpzc{E}I}(d^{I(j)}) + \frac{1}{2} (d^{\mathpzc{E}(j)})^{t}A_{\mathpzc{EE}}(d^{\mathpzc{E}(j)}) - \frac{1}{2}(d^{\mathpzc{E}(j)})^{t}A_{\mathpzc{EE}}(b^{\mathpzc{E}(j)}) \nonumber \\
         &  &   - \frac{1}{2}(b^{\mathpzc{E}(j)})^{t}A_{\mathpzc{EE}}(d^{\mathpzc{E}(j)}) + \frac{1}{2}(b^{\mathpzc{E}(j)})^{t}A_{\mathpzc{EE}}(b^{\mathpzc{E}(j)}) + \frac{1}{2}(d^{I(j)})^{t}A_{II}(d^{I(j)}) .
\end{eqnarray}
}

\indent Take the matrix $\overline{A}$ (Eq. \ref{expmu}) having the form
{\footnotesize
\begin{equation}\label{mtxextend}
\overline{A} = \left( 
\begin{array}{c  c  c c} 
A_{XX} & A_{X\mathpzc{E}} & A_{XI} & A_{XL} \\
A_{LX} & A_{L\mathpzc{E}} & A_{LI} & A_{LL} \\
A_{\mathpzc{E}X} & A_{\mathpzc{EE}} & A _{\mathpzc{E}I} & A_{\mathpzc{E}L} \\
A_{IX}  &  A_{I\mathpzc{E}}  & A_{II} & A_{IL} \\
\end{array}
\right) .
\end{equation}
}

Replacing the matrix (\ref{mtxextend}) in the Eq. (\ref{expmu}) and considering that the submatrices $A_{XX}=0$ (by Def. \ref{ingred}) and $A_{LL}=0$ (by condition C3 in Def. \ref{condgraphdecod}) and $A_{XL}=(A_{LX})^{t}=0$ (by condition C4 in Def. \ref{condgraphdecod}) and $A_{X\mathpzc{E}}=(A_{\mathpzc{E}X})^{t}$, $A_{XI}=(A_{IX})^{t}$, $A_{\mathpzc{E}I}=(A_{I\mathpzc{E}})^{t}$, $A_{\mathpzc{E}L}=(A_{L\mathpzc{E}})^{t}$, $A_{IL}=(A_{LI})^{t}$.  After performing the calculations, we obtain
{\footnotesize
\begin{eqnarray}\label{defmu2}
\mu (j) & = & + (d^{X'})^{t}A_{X\mathpzc{E}}(d^{\mathpzc{E}(j)}) + (d^{X'})^{t}A_{XI}(d^{I(j)}) + (d^{\mathpzc{E}(j)})^{t}A_{\mathpzc{E}I}(d^{I(j)}) + (d^{\mathpzc{E}(j)})^{t}A_{\mathpzc{E}L}(d^{L}) \nonumber \\
         &   &+ \frac{1}{2} (d^{\mathpzc{E}(j)})^{t}A_{\mathpzc{EE}}(d^{\mathpzc{E}(j)}) + (b^{I(j)})^{t}A_{IL}(d^{L}) + \frac{1}{2}(d^{I(j)})^{t}A_{II}(d^{I(j)}) .
\end{eqnarray}
}

\indent Substituting Eqs. (\ref{iqftgrafo}) and (\ref{errorstate2b}) in Eq. (\ref{decodgraph}), we have
{\footnotesize
\begin{eqnarray}\label{decgraph2}
\mathbf{\mathcal{R}} (\vert \varphi \rangle ) & = &  \sum_{j=1}^{p^{|Y|}} \left[ \left( \sum_{d^{X}} e^{(\frac{2 \pi i}{p}).[\eta (j)] } . c(d^{X})  \right) . \left( e^{(\frac{2 \pi i}{p})[(s^{\mathpzc{E}(j)})^{t}.(d^{\mathpzc{E}(j)})]} \right)   \right. \nonumber \\
 & & . \left. \left( \sum_{d^{L}}  \sum_{d^{X'}}  e^{-(\frac{2\pi i}{p}) \left[ \mu (j) \right] } \vert d^{L}d^{X'} \rangle  \right) \right]
\end{eqnarray}
}
where $\eta (j)$ and $\mu (j)$ are given respectively by Eqs. (\ref{eta3}) and (\ref{defmu2}). 

\indent After some manipulations in Eq. (\ref{decgraph2}), we obtain the following expression
{\footnotesize
\begin{eqnarray}\label{decgraph3}
\mathbf{\mathcal{R}} (\vert \varphi \rangle ) & = & \sum_{j=1}^{p^{|Y|}}  \left[ \left( e^{(\frac{2 \pi i}{p}) \left\{ - (b^{\mathpzc{E}(j)})^{t}A_{\mathpzc{E}I}(d^{I(j)}) - \frac{1}{2}(d^{\mathpzc{E}(j)})^{t}A_{\mathpzc{EE}}(b^{\mathpzc{E}(j)}) - \frac{1}{2}(b^{\mathpzc{E}(j)})^{t}A_{\mathpzc{EE}}(d^{\mathpzc{E}(j)})  \right. } \right.   \right.  \nonumber \\
&  & \left. ^{ \left.  + \frac{1}{2}(b^{\mathpzc{E}(j)})^{t}A_{\mathpzc{EE}}(b^{\mathpzc{E}(j)}) \right\} } \right) .  \left( e^{(\frac{2 \pi i}{p})[(s^{\mathpzc{E}(j)})^{t}.(d^{\mathpzc{E}(j)})]} \right) . \left( \sum_{d^{X}} e^{(\frac{2 \pi i}{p}).\left[ (d^{X})^{t}A_{X\mathpzc{E}}(d^{\mathpzc{E}(j)}) \right. }    \right.  \nonumber \\
&  & \left. \left. \left.  ^{  \left. - (d^{X})^{t}A_{X\mathpzc{E}}(b^{\mathpzc{E}(j)}) + (d^{X})^{t}A_{XI}(d^{I(j)})  \right] } . c(d^{X}) \right) \left( \sum_{d^{L}}  \sum_{d^{X'}}  e^{-(\frac{2\pi i}{p}) \left[ (d^{X'})^{t}A_{X\mathpzc{E}}(d^{\mathpzc{E}(j)})   \right. }   \right. \right. \right. \nonumber \\
&  & \left. \left. ^{  \left. + (d^{X'})^{t}A_{XI}(d^{I(j)}) + (d^{\mathpzc{E}(j)})^{t}A_{\mathpzc{E}L}(d^{L}) + (b^{I(j)})^{t}A_{IL}(d^{L}) \right] } \vert d^{L}d^{X'} \rangle \right) \right] 
\end{eqnarray}
}

\indent We can rewrite the above expression as follows
{\footnotesize
\begin{eqnarray}\label{decgraph4}
\mathbf{\mathcal{R}} (\vert \varphi \rangle ) & = & \sum_{d^{X}} \left\{ \sum_{j=1}^{p^{|Y|}}  \left[ \sum_{d^{L}} \sum_{d^{X'}} \left( e^{ (\frac{2 \pi i}{p}) \left[ (s^{\mathpzc{E}(j)})^{t}.(d^{\mathpzc{E}(j)}) + (d^{X})^{t}A_{X\mathpzc{E}}(d^{\mathpzc{E}(j)}) + (d^{I(j)})^{t}A_{IX}(d^{X})  \right. } \right.  \right. \right. \nonumber \\
&  & \left. ^{ \left. - (d^{X'})^{t}A_{X\mathpzc{E}}(d^{\mathpzc{E}(j)}) -  (d^{I(j)})^{t}A_{IX}(d^{X'}) - (d^{\mathpzc{E}(j)})^{t}A_{\mathpzc{E}L}(d^{L}) \right] } \right) . \left( e^{(\frac{2 \pi i}{p}) \left[  \frac{1}{2}(b^{\mathpzc{E}(j)})^{t}A_{\mathpzc{EE}}(b^{\mathpzc{E}(j)})  \right.  } \right. \nonumber \\
& &  \left. ^{ \left. - (b^{\mathpzc{E}(j)})^{t}A_{\mathpzc{EE}}(d^{\mathpzc{E}(j)}) - (d^{I(j)})^{t}A_{I\mathpzc{E}}(b^{\mathpzc{E}(j)}) - (d^{X})^{t}A_{X\mathpzc{E}}(b^{\mathpzc{E}(j)}) + (b^{I(j)})^{t}A_{IL}(d^{L})     \right] } \right) \nonumber \\
& & \left. \left. \vert d^{L}d^{X'} \rangle \right]  c(d^{X}) \right\} .
\end{eqnarray}
}
From  the Eq. (\ref{decgraph4}) we can conclude that the first exponential  figures out a phase-flip error, while the second exponential recognizes a bit-flip 
error.  The exponential va\-lues  are passed to the syndrome qubits $L$ in $\vert d^{L}d^{X'} \rangle$.  After that,  it is  possible recover the original states 
by consulting a look up table of syndrome and corresponding errors. This completes the proof of Lemma \ref{graphdecod}. \hspace{2cm} $\square$

\indent An application making use the Lemmas \ref{defcodgraf} and \ref{graphdecod} will be presented in Section \ref{conscodigo}.

\section{A Quantum Erasure-Correcting Code via GHZ States}\label{erasurecode}
In this section we present a generalization of the scheme proposed by Yang et al.\cite{8} for a quantum erasure-correcting code. 

An arbitrary state of $n$ qubits can be written as follows:
{\footnotesize
\begin{equation}\label{nqubit}
\vert v \rangle = \sum_{i=0}^{2^{n}} \lambda_{i} \vert i \rangle,
\end{equation}
}
where $\sum_{i=0}^{2^{n}} |\lambda_{i}|^{2} = 1$; and $\vert i \rangle$ represents a general basis state of $n$ qubits with the integer $i$
corresponding to its binary decomposition. To hide $n$-qubit quantum information, we can use $n$ ancillary qubits to encode the state (\ref{nqubit}) into

{\footnotesize
\begin{equation}\label{nancillary}
\vert \psi \rangle_{L} = \sum_{i=0}^{2^{n}} \lambda_{i} \vert  \psi^{(i)} \rangle_{12\ldots n} \otimes \vert  \psi^{(i)} \rangle_{1'2'\ldots n'} ,
\end{equation}
}
where $\vert  \psi^{(i)} \rangle_{12\ldots n}$ and $\vert  \psi^{(i)} \rangle_{1'2'\ldots n'}$ are the two $n$-qubit GHZ states, respectively, corresponding to the $n$ qubits ($1, 2,\ldots , n $) and the $n$ ancillary qubits ($1', 2',\ldots , n' $), which are given by

{\footnotesize
\begin{eqnarray}\label{GHZstates}
\vert  \psi^{(i)} \rangle_{12\ldots n} & = & \frac{1}{\sqrt{2}} \left[ \vert u_{1}^{(i)} u_{2}^{(i)} \ldots u_{n}^{(i)} \rangle \pm  \vert \hat{u}_{1}^{(i)} \hat{u}_{2}^{(i)} \ldots \hat{u}_{n}^{(i)} \rangle \right], \nonumber \\
\vert  \psi^{(i)} \rangle_{1'2'\ldots n'} & = & \frac{1}{\sqrt{2}} \left[ \vert v_{1'}^{(i)} v_{2'}^{(i)} \ldots v_{n'}^{(i)} \rangle \pm  \vert \hat{v}_{1'}^{(i)} \hat{v}_{2'}^{(i)} \ldots \hat{v}_{n'}^{(i)} \rangle \right]
\end{eqnarray}
}
(here, $\vert  u_{k}^{(i)} \rangle$ and $\vert  \hat{u}_{k}^{(i)} \rangle$ represent two orthogonal states of the qubit k, $\hat{u}_{k}^{(i)} = 1-u_{k}^{(i)}$ and $u_{k}^{(i)} \in \{ 0, 1 \}$; the same notation holds for the two orthogonal states $\vert  v_{k'}^{(i)} \rangle$ and $\vert  \hat{v}_{k'}^{(i)} \rangle$ of the ancillary qubit $k'$).

Since any basis state in (\ref{nqubit}) is encoded into a product of two $n$-qubit GHZ states, it is straightforward to show that for the encoded state (\ref{nancillary}), the density operator of each qubit is given by $\frac{1}{2} (\vert 0 \rangle \langle 0 \vert + \vert 1 \rangle \langle 1 \vert)$. This result means that the $n$-qubit quantum information, originally carried by the n ``message'' qubits, is hidden over each qubit after encoding the state (\ref{nqubit}) into (\ref{nancillary}). 

The encoding can be easily done by using Hadamard gates and CNOT gates. For simplicity, we consider the case when each basis state in (\ref{nqubit}) is encoded into a product of two $n$-qubit GHZ states both taking the same form. The encoding operation is given by 

{\footnotesize
\begin{equation}\label{opencod}
U_{e} = \prod_{i=1}^{n-1} C_{n'i'} \otimes \prod_{i=1}^{n-1} C_{ni} \otimes H_{n'} H_{n} \otimes \prod_{i=1}^{n} C_{ii'} ,
\end{equation}
}
where the $n$ ancillary qubits are initially in the state $\vert 00\ldots 0 \rangle$; $H_{n}$ and $H_{n'}$ are Hadamard transformation operations, respectively, acting on the qubit $n$ and the ancillary qubit $n'$; $C_{ii'}$ is a CNOT operation acting on the qubit $i$ (control bit) and the ancillary qubit $i'$ (target bit); $C_{ni}$ is a CNOT operation acting on the qubit $n$ (control bit) and the qubit $i$ (target bit); and $C_{n'i'}$ is a CNOT operation acting on the ancillary qubit $n'$ (control bit) and the ancillary qubit $i'$ (target bit). 

\indent It should be mentioned that a general theory about quantum data hiding has been proposed.\cite{19} Althoug we treat a special case that a single party cannot gain any information about the state, our main purpose is to wish to present a concrete encoding scheme for hiding n-qubit information over each qubit. This scheme also provides a good illustration of the relationship between quantum data hiding and QECC already noted in Refs. \cite{19} and \cite{20}, since it is straightforward to show that the above encoding is also equivalent to a QECC correcting one erasure. 

\indent Now, we present the decoding and recovery operations. 

\indent When the erasure occurs in one of the $n$ qubits, the decoding operation is performed by
{\footnotesize
\begin{equation}\label{opdecod1}
U_{d} = H_{n'} \otimes \prod_{i=1}^{n-1} C_{n'i'} .
\end{equation}
}
\indent When the erasure occurs in one of the $n$ ancillary qubits, the decoding operation is performed by
{\footnotesize
\begin{equation}\label{opdecod2}
U_{d} = H_{n} \otimes \prod_{i=1}^{n-1} C_{ni} .
\end{equation}
}

\indent For recovery operation, we consider that the erasure occurred in one of $n$ ``message'' qubits.

\indent To erasure on the qubit $j$, where $j\neq n$, we use
{\footnotesize
\begin{equation}\label{oprecov1}
U_{r} = T_{j'n'k} \otimes Z_{n'k}\otimes T_{j'n'k} \otimes \prod_{i=1  (i \neq j)}^{n-1} C_{i'i} \otimes \prod_{i=1 (i \neq j)}^{n} C_{j'i}  ,
\end{equation}  
}
where $k=n-j$, $T_{j'n'k}$ is a Toffoli gate operation,\cite{11} and $Z_{n'k}$ is a controlled Pauli $\sigma_{Z}$ operation. A Toffoli gate $T_{jnk}$ has two control bits corresponding to the first two subscripts ($j,n$), and the target bit $k$. When the two control bits are in the state $\vert 11 \rangle$, the state of the target bit will change, following $\vert 0 \rangle \rightarrow \vert 1 \rangle $ and $\vert 1 \rangle \rightarrow \vert 0 \rangle$, while when the two control bits are in the state $\vert 00 \rangle$, $\vert 01 \rangle$ or $\vert 10 \rangle$, the state of the target bit will be invariant. The controlled Pauli $\sigma_{Z}$ operation $Z_{nk}$ has the control bit $n$ and  target bit $k$, which sends the state of the target bit $\vert 0 \rangle \rightarrow \vert 0 \rangle$ and $\vert 1 \rangle \rightarrow - \vert 1 \rangle$ when the control bit is in the state $\vert 1 \rangle$; otherwise, when the control bit is in the $\vert 0 \rangle$, the state of the target bit will not change. 

\indent To erasure on the qubit $j$, where $j=n$, we use
{\footnotesize
\begin{equation}\label{oprecov1}
U_{r} = Z_{j'k}\otimes \prod_{i=1 (i \neq j)}^{n-1} C_{i'i} .
\end{equation} 
}
where here $k=n-1$.

\indent The recovery operator for the case where the erasure occurs in one of $n$ ancillary qubits is similar to these.

\indent Taking into account the price which we will probably have to pay in deter\-mi\-ning the error position, the fact that we have to know which qubit goes ``bad''. But again, it is compensated for by the fact that we need a smaller number of ancillary qubits to construct a quantum erasure-correcting code, for example, only one ancillary qubit is required for one qubit on average as far as the present code. Also, as shown above, since the ``damaged'' particle is not involved in the recovery operations, the present code can still work in the case when the interaction with environment leads to the leakage of a qubit out of the qubit space. 

\indent So, we have presented a $2n$-qubit code for protecting $n$-qubit quantum information against one erasure. There are already experimental studies that used similar schemes to that presented above to protect a logical qubit from loss of a physical qubit, see Ref. \cite{21}. What proves in-principle the experimental feasibility of this scheme to protect against the occurrence of erasure.  The encoding, decoding and error recovery operations, as shown here, are relatively straightforward. A special feature of the error recovery method is that no extra ancillary qubits and no measurement are required.

\section{A Concatenated Quantum Code}\label{conscodigo}
In this section we will describe how two codes may be concatenated to form a large code, based on Ref. \cite{22}. After that, we  present a scheme of concatenated code for protecting against both occurrence of quantum computational error and a quantum erasure. Finally, we will illustrate this scheme of concatenated quantum code with a simple example which will be able to protecting one qubit against both occurrences of a quantum computational error and one quantum erasure. 

\indent Let the two codes bean $M$-qubit code $\mathbf{C}^{out}=(\mathbf{E}^{out}, \mathbf{D}^{out})$, called the outer code, and an $N$-qubit code $\mathbf{C}^{in}=(\mathbf{E}^{in}, \mathbf{D}^{in})$, called the inner code. A logical qubit $\rho_{0}$ is encoded first using the outer code $\mathbf{C}^{out}$, yielding the $M$-qubit state $\mathbf{E}^{out}[\rho_{0}]$. Each of these qubits is then encoded by the inner code; i.e., the map $\mathbf{E}^{in} \otimes \ldots \otimes \mathbf{E}^{in} = (\mathbf{E}^{in})^{\otimes M}$ acts on $\mathbf{E}^{out}[\rho_{0}]$. The composition of these encodings forms the encoding map for the concatenated code:

{\footnotesize
\begin{equation}\label{encodconc}
\tilde{\mathbf{E}} =  (\mathbf{E}^{in})^{\otimes M} \circ \mathbf{E}^{out} .
\end{equation}
}
The $M$ sections of the register encoding each of the $M$ qubits in $\mathbf{E}^{out} [\rho_{0}]$ are called blocks; each block constains $N$ qubits. After the encoding, a noise process $\tilde{\mathcal{N}}$ acts on the entire $MN$-qubit register.

A simple error-correction scheme (and one that seems reasonable for use in a scalable architecture) coherently corrects each of the code blocks based on the inner code, and then corrects the entire register based on the outer code. That is, the decoding map for the concatenated code is given by

{\footnotesize
\begin{equation}\label{decodconc}
\tilde{\mathbf{D}}= \mathbf{D}^{out} \circ \mathbf{D}^{in \otimes M} .
\end{equation}
}
We denote the concatenated code (with this correction scheme) by $\mathbf{C}^{out} (\mathbf{C}^{in})=(\tilde{\mathbf{E}}, \tilde{\mathbf{D}})$; note that $\mathbf{C}^{out} (\mathbf{C}^{in})$ is an $MN$-qubit code.

Suppose that we have computed the effective channel $\mathfrak{C}_{ef}$ due to a code $\mathbf{C}^{in}=(\mathbf{E}^{in}, \mathbf{D}^{in})$ with some noise dynamics $\mathcal{N}$, and wish to consider the effective channel $\tilde{\mathfrak{C}}_{ef}$ resulting from the concatenated code $\mathbf{C}^{out} (\mathbf{C}^{in})$. We assume that each $N$-bit block in the register evolves according to the noise dynamics $\mathcal{N}$ and no cross-block correlations are introduced, i.e., that the evolution operator on the $MN$-bit register is

{\footnotesize
\begin{equation}\label{generalnoise}
\tilde{\mathcal{N}} = \mathcal{N} \otimes \mathcal{N} \otimes \ldots \otimes \mathcal{N} = \mathcal{N}^{\otimes M} .
\end{equation}
}
By definition, we have $\tilde{\mathfrak{C}}_{ef} = \tilde{\mathbf{D}} \circ \tilde{\mathcal{N}} \circ \tilde{\mathbf{E}}$. Substituting Eqs. (\ref{encodconc}), (\ref{decodconc}), and (\ref{generalnoise}) into this expression yields
{\footnotesize
\begin{eqnarray}
\tilde{\mathfrak{C}}_{ef} & = & \mathbf{D}^{out} \circ \mathbf{D}^{in \otimes M} \circ \mathcal{N}^{\otimes M} \circ \mathbf{E}^{in \otimes M} \circ \mathbf{E}^{out} \nonumber \\
                                  & = & \mathbf{D}^{out} \circ (\mathbf{D}^{in} \circ \mathcal{N} \circ \mathbf{E}^{in})^{\otimes M} \circ \mathbf{E}^{out} \nonumber \\
                                  & = & \mathbf{D}^{out} \circ \mathfrak{C}_{ef}^{\otimes M} \circ \mathbf{E}^{out},
\end{eqnarray}
}
where we have used $\mathfrak{C}_{ef}= \mathbf{D}^{in} \circ \mathcal{N} \circ \mathbf{E}^{in}$. This result makes sense: each of the $M$ blocks of $N$ bits represents a single logical qubit encoded in $\mathbf{C}^{in}$, and as the block has dynamics $\mathcal{N}$, this logical qubit's evolution will be described by $\mathfrak{C}_{ef}$.

\indent Thinking of correct changes caused by the environment, it is straightforward to think first in correcting situations in which the position of the erroneous qubits is known (erasures) and then to correct situations in which the position of the erroneous qubits are not known (computational errors). Considering this, for the concatenated code presented here, the internal code will protect the information against the occurrence of erasure, while the external code will protect the information against the occurrence of computational errors. 

\indent This concatenation scheme will lead to external code the quantum graph code, which will make the protection against computacional quantum errors. As internal code, will be used the quantum erasure-correcting code via GHZ states to achieve protection against one erasure.

\indent To illustrate how the protection of information can be achieved by this scheme of concatenation,  we will consider a simple example, which has as external code the quantum code [[5,1,3]] obtained via a 3-regular graph, which will protect the information against a computational error. As internal code, to protect against one erasure, we use operators which deal with GHZ states of five-qubits, such operators have been constructed as seen in Section \ref{erasurecode}.

\indent The code [[5,1,3]], the smallest Hamming code, was independently discovered by Bennet et al.\cite{23} and Laflamme at al.\cite{24}. A new proof of the existence of this quantum code was given by Schlingemann and Werner \cite{1} via graph codes, for that they made use of the wheel graph $W_{6}$. In the present work, we will use a 3-regular graph.

\begin{figure}[h]
\centering
\includegraphics[scale=0.2]{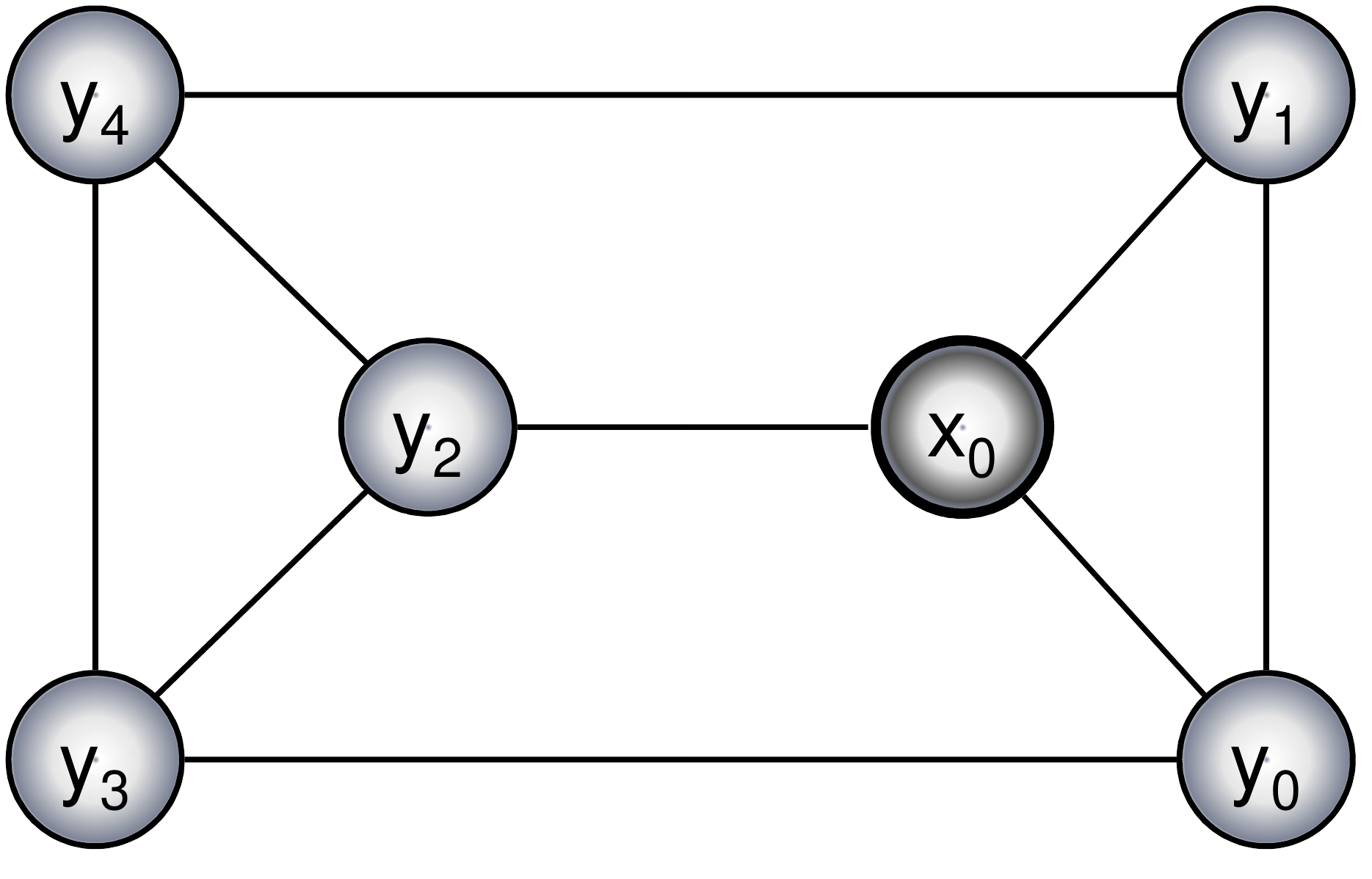}
\caption{\footnotesize A geometric representation of a graph for the code [[5, 1, 3]].}
\label{figure:hexacodigo1}
\end{figure}

\indent For the code [[5,1,3]] we have $n=5$, $k = 1$ and $d = 3$. Thus, $ |X| = 1$ and $| Y | = 5$ and therefore the 3-regular graph considered has six vertices, as can be seen in Figure \ref{figure:hexacodigo1}. The graph in focus generates a QECC, as shown by Santos.\cite{25} 

\indent Considering the field $\mathbb{Z}_{2}$, this graph has matrix 
{\footnotesize
\begin{equation}\label{mtxgama}
A = \left( 
\begin{array}{c | c} 
A_{XX} & A_{XY} \\ \hline
A_{YX} & A_{YY} \\
\end{array}
\right) 
= 
\begin{array}{c}
x_{0} \\
y_{0} \\
y_{1} \\
y_{2} \\
y_{3} \\
y_{4} \\ 
\end{array}
\stackrel {
\begin{array}{l l l l l l}
 \ \ x_{0}   &   y_{0}   &   y_{1}   &   y_{2}   &   y_{3}   &   y_{4} \\ 
\end{array} }
{\left(
\begin{array}{l | r r r r r}
        0    & \  1  &  \ 1   &  \  1   & \ 0   & \  0   \\ \hline
       1    &  \ 0  &  \  1   &  \  0    & \ 1   & \  0  \\
       1    &  \ 1  &  \ 0   &  \ 0    & \ 0   &  \ 1  \\
       1    &  \ 0  & \  0   & \  0    & \  1  &  \ 1   \\
       0    &  \ 1  &  \ 0   &  \ 1    & \ 0   & \  1   \\
       0    &  \ 0  &  \  1  &  \  1   & \  1  & \  0  \\
\end{array}
\right). }
\end{equation}
}

\indent Consider the vertice $x_{0}$ as input vertice and the remaining five vertices as output, that is, $y_{0}, y_{1}, y_{2}, y_{3}, y_{4} \in Y$ (see Figure \ref{figure:hexacodigo1}). Then 

{\footnotesize
$$
d^{X} = \left( x_{0} \right); \ \  d^{Y} = \left( \begin{array}{c} y_{0} \\ y_{1} \\ y_{2} \\ y_{3} \\ y_{4} \end{array} \right)  .
$$
}

Therefore, according to Lemma \ref{defcodgraf}, the quantum code for this graph is given as follows (To simplify the notation, the normalization factors are omitted here and in the rest of this work): 

{\footnotesize
$$
\vert v \rangle = \sum_{x_{0} \in \mathbb{F}_{p}^{X}} c(x_{0}) \vert x_{0} \rangle , \mbox{em que} \ c(x_{0}) \in \mathbb{C}
$$
}
and
{\footnotesize
\begin{equation}\label{eqcod3r}
f(\vert v \rangle ) = \sum_{( y_{0}, y_{1}, y_{2}, y_{3}, y_{4}) \in \mathbb{F}_{p}^{Y}} \sum_{x_{0} \in \mathbb{F}_{p}^{X}} e^{\frac{2 \pi i}{p} \left( \theta \right) } c (x_{0} ) \vert y_{0} y_{1} y_{2} y_{3} y_{4} \rangle .
\end{equation}
}
where $\theta =x_{0} y_{0} + x_{0} y_{1} + x_{0} y_{2} + y_{0} y_{1} + y_{0} y_{3} + y_{1} y_{4} + y_{2} y_{3} + y_{2} y_{4} + y_{3} y_{4}$. 

\indent To illustrate the obtention of the code graph, we consider the elements in $\mathbb{Z}_{2} =\{ 0, 1 \}$. In this form, 

{\footnotesize
$$
\vert v \rangle = c(0) \vert 0 \rangle + c(1) \vert 1 \rangle .
$$
}

Substituting the values of $\mathbb{Z}_{2} =\{ 0, 1 \}$ in the Eq. (\ref{eqcod3r}) above and performing the calculations, we obtain the encoded state
{\footnotesize
\begin{eqnarray}\label{estadografo}
f(\vert v \rangle ) = \vert \psi \rangle_{12345} & = & \lambda_{0}  \vert 00000  \rangle  + \lambda_{1}  \vert 00001  \rangle  + \lambda_{2}  \vert 00010  \rangle + \lambda_{3}  \vert 00011  \rangle  + \lambda_{4}  \vert 00100  \rangle  + \lambda_{5}  \vert 00101  \rangle   \nonumber \\
& &  + \lambda_{6}  \vert 00110  \rangle  + \lambda_{7}  \vert 00111  \rangle  + \lambda_{8}  \vert 01000  \rangle + \lambda_{9}  \vert 01001  \rangle  + \lambda_{10} \vert 01010  \rangle  + \lambda_{11}  \vert 01011  \rangle  \nonumber \\
& &  + \lambda_{12} \vert 01100  \rangle +  \lambda_{13} \vert 01101  \rangle  + \lambda_{14} \vert 01110  \rangle  + \lambda_{15} \vert 01111  \rangle  + \lambda_{16} \vert 10000  \rangle  + \lambda_{17} \vert 10001  \rangle  \nonumber \\
& &  + \lambda_{18} \vert 10010  \rangle + \lambda_{19} \vert 10011  \rangle  + \lambda_{20} \vert 10100  \rangle + \lambda_{21} \vert 10101  \rangle + \lambda_{22} \vert 10110  \rangle  + \lambda_{23} \vert 10111  \rangle  \nonumber \\
 & & + \lambda_{24} \vert 11000  \rangle  + \lambda_{25} \vert 11001  \rangle + \lambda_{26} \vert 11010  \rangle + \lambda_{27} \vert 11011  \rangle  + \lambda_{28}  \vert 11100  \rangle  + \lambda_{29} \vert 11101  \rangle   \nonumber \\
 & &  + \lambda_{30} \vert 11110  \rangle  + \lambda_{31}  \vert 11111  \rangle , 
\end{eqnarray}
}
where
{\footnotesize
\begin{eqnarray}
\lambda_{0} = c(0)  + c(1);   & \lambda_{1} = c(0)  + c(1);  \ \    \lambda_{2} = c(0)  + c(1);    & \lambda_{3} = - c(0)  - c(1) ;    \nonumber \\
\lambda_{4} = c(0)  - c(1) ;   & \lambda_{5} = - c(0)  + c(1) ;  \ \  \lambda_{6} = - c(0)  + c(1);  & \lambda_{7} = - c(0)  + c(1);   \nonumber \\
\lambda_{8} = c(0)  - c(1);    & \lambda_{9} = - c(0)  + c(1);   \ \ \lambda_{10} = c(0)  - c(1) ;   & \lambda_{11} = c(0)  - c(1);  \nonumber \\
\lambda_{12} = c(0)  + c(1);  & \lambda_{13} = c(0)  + c(1) ; \ \  \lambda_{14} = - c(0)  - c(1) ; &  \lambda_{15} = c(0)  + c(1);   \nonumber \\
\lambda_{16} = c(0)  - c(1);  & \lambda_{17} = c(0)  - c(1) ; \ \  \lambda_{18} =  - c(0)  + c(1);  & \lambda_{19} = c(0)  - c(1);   \nonumber \\
\lambda_{20} = c(0)  + c(1);  & \lambda_{21} = - c(0)  - c(1); \ \ \lambda_{22} = c(0)  + c(1);   & \lambda_{23} = c(0) + c(1);   \nonumber \\
\lambda_{24} = - c(0)  - c(1);    & \lambda_{25} = c(0)  + c(1) ; \ \  \lambda_{26} = c(0)  + c(1); &  \lambda_{27} = c(0)  + c(1) ;    \nonumber \\
\lambda_{28} = - c(0)  + c(1); & \lambda_{29} = - c(0) + c(1); \ \ \lambda_{30} = - c(0)  + c(1);   & \lambda_{31} = c(0)  - c(1)  . 
\end{eqnarray} 
}

\indent The encoding performed above will protect a qubit against the occurrence of a computational error. 

\indent The scheme that we will use from now on will use GHZ states of five-qubits to achieve the protection of the encoded state $\vert \psi \rangle_{12345}$ against an erasure, for this operators were built based on the Section \ref{erasurecode}.

\indent The produced code requires five ancillary qubits (giving a total of ten-qubits) for the protection of five-qubits against the occurrence of an erasure.

\indent Using five ancillary qbits $(1'2'3'4'5')$, we can encode the state $\vert \psi \rangle_{12345}$, in a way that 
{\footnotesize
\begin{equation}\label{opercod1}
\mathcal{U}_{e} (\vert \psi \rangle_{12345} \otimes \vert 00000 \rangle_{1'2'3'4'5'})= \vert \varphi \rangle ,
\end{equation}
}
where 
{\footnotesize
\begin{equation}\label{opercod2}
\mathcal{U}_{e} = C_{5'4'}C_{5'3'}C_{5'2'}C_{5'1'}C_{54}C_{53}C_{52}C_{51}H_{5'}H_{5}C_{55'}C_{44'}C_{33'}C_{22'}C_{11'} .
\end{equation}
}

\indent Note that the five ancillary qbits $1', 2', 3', 4', 5 '$ are initially in the state $\vert 00000 \rangle$. Throughout this work, every joint operation, as outlined above, will follow the sequence from right to left.

The encoding performed by the operator (\ref{opercod2}) makes use of quantum CNOT (controlled-NOT) operations, where the first subscript of $C_{ij}$  refers to the control bit, and the second, to target bit, and by Hadarmard transformatioon $H_{i}$ in qubit $i$, which sends $\vert 0 \rangle \rightarrow (\vert 0 \rangle + \vert 1 \rangle)$ and $\vert 1 \rangle \rightarrow (\vert 0 \rangle - \vert 1 \rangle)$. So, after this encoding, we obtain 

{\scriptsize
\begin{eqnarray}
\vert \varphi \rangle & = & \lambda_{0}  \vert 0  \rangle_{L}  + \lambda_{1}  \vert 1  \rangle_{L}  + \lambda_{2}  \vert 2  \rangle_{L} + \lambda_{3}  \vert 3  \rangle_{L}  + \lambda_{4}  \vert 4  \rangle_{L}  + \lambda_{5}  \vert 5  \rangle_{L}  + \lambda_{6}  \vert 6  \rangle_{L}  + \lambda_{7}  \vert 7  \rangle_{L}  + \lambda_{8}  \vert 8  \rangle_{L} +  \lambda_{9}  \vert 9  \rangle_{L}   \nonumber \\
& &  + \lambda_{10} \vert 10  \rangle_{L}  + \lambda_{11}  \vert 11  \rangle_{L}  + \lambda_{12} \vert 12  \rangle_{L} +  \lambda_{13} \vert 13  \rangle_{L}  + \lambda_{14} \vert 14  \rangle_{L}  + \lambda_{15} \vert 15  \rangle_{L}  + \lambda_{16} \vert 16  \rangle_{L}  + \lambda_{17} \vert 17  \rangle_{L}  \nonumber \\
& &  + \lambda_{18} \vert 18  \rangle_{L} + \lambda_{19} \vert 19  \rangle_{L}  + \lambda_{20} \vert 20  \rangle_{L} + \lambda_{21} \vert 21  \rangle_{L} + \lambda_{22} \vert 22  \rangle_{L}  + \lambda_{23} \vert 23  \rangle_{L} + \lambda_{24} \vert 24  \rangle_{L}  + \lambda_{25} \vert 25  \rangle_{L}   \nonumber \\
 & & + \lambda_{26} \vert 26  \rangle_{L} +  \lambda_{27} \vert 27  \rangle_{L}  + \lambda_{28}  \vert 28  \rangle_{L}  + \lambda_{29} \vert 29  \rangle_{L}  + \lambda_{30} \vert 30  \rangle_{L}  + \lambda_{31}  \vert 31  \rangle_{L} , 
\end{eqnarray} 
}
where the thirty-two states are:
{\footnotesize
\begin{eqnarray}\label{redund1}
\vert 0 \rangle_{L} & = & (\vert 00000 \rangle + \vert 11111 \rangle)_{12345} \otimes (\vert 00000 \rangle + \vert 11111 \rangle)_{1'2'3'4'5'} , \nonumber \\
\vert 1 \rangle_{L} & = & (\vert 00000 \rangle - \vert 11111 \rangle)_{12345} \otimes (\vert 00000 \rangle - \vert 11111 \rangle )_{1'2'3'4'5'}, \nonumber \\
\vert 2 \rangle_{L} & = & (\vert 00010 \rangle + \vert 11101 \rangle)_{12345} \otimes (\vert 00010 \rangle + \vert 11101 \rangle)_{1'2'3'4'5'}, \nonumber \\
\vert 3 \rangle_{L} & = & (\vert 00010 \rangle - \vert 11101 \rangle)_{12345} \otimes (\vert 00010 \rangle - \vert 11101 \rangle)_{1'2'3'4'5'}, \nonumber \\
\vert 4 \rangle_{L} & = & (\vert 00100 \rangle + \vert 11011 \rangle)_{12345} \otimes (\vert 00100 \rangle + \vert 11011 \rangle)_{1'2'3'4'5'}, \nonumber \\
\vert 5 \rangle_{L} & = & (\vert 00100 \rangle - 11011 \rangle)_{12345} \otimes (\vert 00100 \rangle - \vert 11011 \rangle)_{1'2'3'4'5'}, \nonumber \\
\vert 6 \rangle_{L} & = & (\vert 00110 \rangle + \vert 11001 \rangle)_{12345} \otimes (\vert 00110 \rangle + \vert 11001 \rangle)_{1'2'3'4'5'} ,\nonumber \\
\vert 7 \rangle_{L} & = & (\vert 00110 \rangle - \vert 11001 \rangle)_{12345} \otimes (\vert 00110 \rangle - \vert 11001 \rangle)_{1'2'3'4'5'},  \nonumber \\
\vert 8 \rangle_{L} & = & (\vert 01000 \rangle + \vert 10111 \rangle)_{12345} \otimes (\vert 01000 \rangle + \vert 10111 \rangle)_{1'2'3'4'5'}, \nonumber \\
\vert 9 \rangle_{L} & = & (\vert 01000 \rangle - \vert 10111 \rangle)_{12345} \otimes (\vert 01000 \rangle - \vert 10111 \rangle)_{1'2'3'4'5'} ,\nonumber \\
\vert 10 \rangle_{L} & = & (\vert 01010 \rangle + \vert 10101 \rangle)_{12345} \otimes (\vert 01010 \rangle + \vert 10101 \rangle)_{1'2'3'4'5'}, \nonumber \\
\vert 11 \rangle_{L} & = & (\vert 01010 \rangle - \vert 10101 \rangle)_{12345} \otimes (\vert 01010 \rangle - \vert 10101 \rangle)_{1'2'3'4'5'}, \nonumber \\
\vert 12 \rangle_{L} & = & (\vert 01100 \rangle + \vert 10011 \rangle)_{12345} \otimes (\vert 01100 \rangle + \vert 10011 \rangle)_{1'2'3'4'5'}, \nonumber \\
\vert 13 \rangle_{L} & = & (\vert 01100 \rangle - \vert 10011 \rangle)_{12345} \otimes (\vert 01100 \rangle - \vert 10011 \rangle)_{1'2'3'4'5'}, \nonumber \\
\vert 14 \rangle_{L} & = & (\vert 01110 \rangle + \vert 10001 \rangle)_{12345} \otimes (\vert 01110 \rangle + \vert 10001 \rangle)_{1'2'3'4'5'}, \nonumber \\
\vert 15 \rangle_{L} & = & (\vert 01110 \rangle - \vert 10001 \rangle)_{12345} \otimes (\vert 01110 \rangle - \vert 10001 \rangle)_{1'2'3'4'5'}, \nonumber \\
\vert 16  \rangle_{L} & = & (\vert 10000 \rangle + \vert 01111 \rangle)_{12345} \otimes (\vert 10000 \rangle + \vert 01111 \rangle)_{1'2'3'4'5'}, \nonumber \\
\vert 17  \rangle_{L} & = & (\vert 10000 \rangle - \vert 01111 \rangle)_{12345} \otimes (\vert 10000 \rangle - \vert 01111 \rangle)_{1'2'3'4'5'}, \nonumber \\
\vert 18  \rangle_{L} & = & (\vert 10010 \rangle + \vert 01101 \rangle)_{12345} \otimes (\vert 10010 \rangle + \vert 01101 \rangle)_{1'2'3'4'5'}, \nonumber \\
\vert 19  \rangle_{L} & = & (\vert 10010 \rangle - \vert 01101 \rangle)_{12345} \otimes (\vert 10010 \rangle - \vert 01101 \rangle)_{1'2'3'4'5'}, \nonumber 
\end{eqnarray}

\begin{eqnarray}
\vert 20  \rangle_{L} & = & (\vert 10100 \rangle + \vert 01011 \rangle)_{12345} \otimes (\vert 10100 \rangle + \vert 01011 \rangle)_{1'2'3'4'5'}, \nonumber \\
\vert 21  \rangle_{L} & = & (\vert 10100 \rangle - \vert 01011 \rangle)_{12345} \otimes (\vert 10100 \rangle - \vert 01011 \rangle)_{1'2'3'4'5'}, \nonumber \\
\vert 22  \rangle_{L} & = & (\vert 10110 \rangle + \vert 01001 \rangle)_{12345} \otimes (\vert 10110 \rangle + \vert 01001 \rangle)_{1'2'3'4'5'},\nonumber \\
\vert 23  \rangle_{L} & = & (\vert 10110 \rangle - \vert 01001 \rangle)_{12345} \otimes (\vert 10110 \rangle - \vert 01001 \rangle)_{1'2'3'4'5'},\nonumber \\
\vert 24  \rangle_{L} & = & (\vert 11000 \rangle + \vert 00111 \rangle)_{12345} \otimes (\vert 11000 \rangle + \vert 00111 \rangle)_{1'2'3'4'5'}, \nonumber \\
\vert 25  \rangle_{L} & = & (\vert 11000 \rangle - \vert 00111 \rangle)_{12345} \otimes (\vert 11000 \rangle - \vert 00111 \rangle)_{1'2'3'4'5'}, \nonumber \\
\vert 26  \rangle_{L} & = & (\vert 11010 \rangle + \vert 00101 \rangle)_{12345} \otimes (\vert 11010 \rangle + \vert 00101 \rangle)_{1'2'3'4'5'}, \nonumber \\
\vert 27  \rangle_{L} & = & (\vert 11010 \rangle - \vert 00101 \rangle)_{12345} \otimes (\vert 11010 \rangle - \vert 00101 \rangle)_{1'2'3'4'5'}, \nonumber \\
\vert 28  \rangle_{L} & = & (\vert 11100 \rangle + \vert 00011 \rangle)_{12345} \otimes (\vert 11100 \rangle + \vert 00011 \rangle)_{1'2'3'4'5'},\nonumber \\
\vert 29  \rangle_{L} & = & (\vert 11100 \rangle - \vert 00011 \rangle)_{12345} \otimes (\vert 11100 \rangle - \vert 00011 \rangle)_{1'2'3'4'5'}, \nonumber \\
\vert 30 \rangle_{L} & = & (\vert 11110 \rangle + \vert 00001 \rangle)_{12345} \otimes (\vert 11110 \rangle + \vert 00001 \rangle)_{1'2'3'4'5'}, \nonumber \\
\vert 31 \rangle_{L} & = & (\vert 11110 \rangle - \vert 00001 \rangle)_{12345} \otimes (\vert 11110 \rangle - \vert 00001 \rangle)_{1'2'3'4'5'}. 
\end{eqnarray}
}

\indent  Note that for the state $\vert \varphi \rangle$ the states of the base keep on being 32 ($2^{5}$), however each logical state is a product of twin GHZ state of five-qubits. It should be also noted that for all the logical states the left part of the product corresponds to the five-qubits of the $\vert \psi \rangle_{12345}$, while the right part of the product corresponds to the five ancillary qubits, and the sequence of the arrangement of the ten-qubits is $1,2,3,4,5,1',2',3',4',5'$ from left to right.

\indent This completes the encoding to protect one qubit of information against both a quantum computational error and a quantum erasure.

\indent After these encodings, the encoded information is submitted to the quantum channel and it is subject to occurrences that can be caused by the environment, as already mentioned previously in the beginning of this work.

Consider that qubit 1 suffers changes caused by the environment. The fact that $\vert 0 \rangle$ and $\vert 1 \rangle$ form a base for the qubit 1, we need only to know what happened in those two states. In general, the process of changes brought about by the environment must be

{\footnotesize
\begin{eqnarray}\label{descqbit}
\vert e_{0} \rangle \vert 0 \rangle & \rightarrow  & \vert \epsilon_{0} \rangle \vert 0 \rangle + \vert \epsilon_{1} \rangle \vert 1 \rangle , \nonumber \\
\\
\vert e_{0} \rangle \vert 1 \rangle &  \rightarrow & \vert \epsilon_{0}' \rangle \vert 0 \rangle + \vert \epsilon_{1}' \rangle \vert 1 \rangle , \nonumber
\end{eqnarray}
}
where $\vert \epsilon_{0} \rangle, \vert \epsilon_{1} \rangle, \vert \epsilon_{0}' \rangle$ e $\vert \epsilon_{1}' \rangle$ are appropriate states of the environment, not necessarily orthogonal or normalized and $\vert e_{0} \rangle$ is the initial state of the environment.

\indent Explicity, consider the situation where the encoded state $\vert \varphi \rangle$ suffers action caused by the environment and cause, for example, erasure in qubit 1 and bit-flip in qubit $1'$. Thus, after these alterations the encoded state will be as follows 

{\scriptsize
\begin{eqnarray}
\vert e_{0} \rangle \otimes \vert \varphi \rangle & = & \lambda_{0}  \vert \hat{\bf 0}  \rangle_{L}  + \lambda_{1}  \vert \hat{\bf 1}  \rangle_{L}  + \lambda_{2}  \vert \hat{\bf 2}  \rangle_{L} + \lambda_{3}  \vert \hat{\bf 3}  \rangle_{L}  + \lambda_{4}  \vert \hat{\bf 4}  \rangle_{L}  + \lambda_{5}  \vert \hat{\bf 5}  \rangle_{L} + \lambda_{6}  \vert \hat{\bf 6}  \rangle_{L}  + \lambda_{7}  \vert \hat{\bf 7}  \rangle_{L}  + \lambda_{8}  \vert \hat{\bf 8}  \rangle_{L} +  \lambda_{9}  \vert \hat{\bf 9}  \rangle_{L}   \nonumber  \nonumber \\
& &  + \lambda_{10} \vert \hat{\bf 10}  \rangle_{L}  + \lambda_{11}  \vert \hat{\bf 11}  \rangle_{L} + \lambda_{12} \vert \hat{\bf 12}  \rangle_{L} +  \lambda_{13} \vert \hat{\bf 13}  \rangle_{L}  + \lambda_{14} \vert \hat{\bf 14}  \rangle_{L} + \lambda_{15} \vert \hat{\bf 15}  \rangle_{L}  + \lambda_{16} \vert \hat{\bf 16}  \rangle_{L}  + \lambda_{17} \vert \hat{\bf 17} \rangle_{L} \nonumber \\
& & + \lambda_{18} \vert \hat{\bf 18}  \rangle_{L} + \lambda_{19} \vert \hat{\bf 19}  \rangle_{L}  + \lambda_{20} \vert \hat{\bf 20}  \rangle_{L} + \lambda_{21} \vert \hat{\bf 21}  \rangle_{L} + \lambda_{22} \vert \hat{\bf 22}  \rangle_{L} + \lambda_{23} \vert \hat{\bf 23}  \rangle_{L}  + \lambda_{24} \vert \hat{\bf 24}  \rangle_{L}  + \lambda_{25} \vert \hat{\bf 25}  \rangle_{L}   \nonumber \\
 & & + \lambda_{26} \vert \hat{\bf 26}  \rangle_{L} + \lambda_{27} \vert \hat{\bf 27}  \rangle_{L}  + \lambda_{28}  \vert \hat{\bf 28}  \rangle_{L} + \lambda_{\bf 29} \vert \hat{\bf 29}  \rangle_{L}  + \lambda_{30} \vert \hat{\bf 30}  \rangle_{L} + \lambda_{31}  \vert \hat{\bf 31}  \rangle_{L}, \nonumber 
\end{eqnarray} 
}
where
{\footnotesize
\begin{eqnarray}\label{redund2}
\vert \hat{\bf 0} \rangle_{L} & = & (\vert \tilde{0}0000 \rangle + \vert \tilde{1}1111 \rangle)_{12345} \otimes (\vert \b{1}0000 \rangle + \vert \b{0}1111 \rangle)_{1'2'3'4'5'}, \nonumber \\
\vert \hat{\bf 1} \rangle_{L} & = & (\vert \tilde{0}0000 \rangle - \vert \tilde{1}1111 \rangle)_{12345} \otimes (\vert \b{1}0000 \rangle - \vert \b{0}1111 \rangle)_{1'2'3'4'5'}, \nonumber \\
\vert \hat{\bf 2} \rangle_{L} & = & (\vert \tilde{0}0010 \rangle + \vert \tilde{1}1101 \rangle)_{12345} \otimes (\vert \b{1}0010 \rangle + \vert \b{0}1101 \rangle)_{1'2'3'4'5'}, \nonumber \\
\vert \hat{\bf 3} \rangle_{L} & = & (\vert \tilde{0}0010 \rangle - \vert \tilde{1}1101 \rangle)_{12345} \otimes (\vert \b{1}0010 \rangle - \vert \b{0}1101 \rangle)_{1'2'3'4'5'}, \nonumber \\
\vdots & \vdots & \vdots \nonumber \\
\vert \hat{\bf 28}  \rangle_{L} & = & (\vert \tilde{1}1100 \rangle + \vert \tilde{0}0011 \rangle)_{12345} \otimes (\vert \b{0}1100 \rangle + \vert \b{1}0011 \rangle)_{1'2'3'4'5'}, \nonumber \\
\vert \hat{\bf 29}  \rangle_{L} & = & (\vert \tilde{1}1100 \rangle - \vert \tilde{0}0011 \rangle)_{12345} \otimes (\vert \b{0}1100 \rangle - \vert \b{1}0011 \rangle)_{1'2'3'4'5'}, \nonumber \\
\vert \hat{\bf 30} \rangle_{L} & = & (\vert \tilde{1}1110 \rangle + \vert \tilde{0}0001 \rangle)_{12345} \otimes (\vert \b{0}1110 \rangle + \vert \b{1}0001 \rangle)_{1'2'3'4'5'}, \nonumber \\
\vert \hat{\bf 31} \rangle_{L} & = & (\vert \tilde{1}1110 \rangle - \vert \tilde{0}0001 \rangle)_{12345} \otimes (\vert \b{0}1110 \rangle - \vert \b{1}0001 \rangle)_{1'2'3'4'5'}, 
\end{eqnarray}
}

\indent Comparing the logical states of (\ref {redund2}) with the ones of (\ref {redund1}), one can notice that for each qubit with erasure (with a tilde on top) of the logical state of (\ref {redund2}), the right part of the product, which matches the encoding of the five ancillary qubits, is not affected by the action of the erasure, however it also suffers an alteration caused by the environment that is the bit-flip (indicated by a dash at the bottom). 

\indent Considering that the occurrence of erasure is by definition an action that in some form is both detectable and locatable, it is natural to think first about correcting the alteration whose occurrence was announced and then see if there were a computational error. Because of this, the first decoding to be performed is the occurrence of erasure. 

First, it performs a unitary transformation in the five ancillary qubits, since in this part an erasure was not detected. This operation is considered as the partial decoding operation (since the qubits 1,2,3,4 and 5 are not involved in the operation of decoding). The decoding operation is as follows 

{\footnotesize
\begin{equation}\label{operdec1}
\mathcal{U}_{d} = H_{5'}C_{5'4'}C_{5'3'}C_{5'2'}C_{5'1'}.
\end{equation}
}

\indent After decoding, we obtain

{\footnotesize
\begin{eqnarray}\label{redund3}
\vert \hat{\bf 0} \rangle_{L} & = & (\vert \tilde{0}0000 \rangle + \vert \tilde{1}1111 \rangle)_{12345} \otimes \vert \b{1}0000 \rangle_{1'2'3'4'5'} , \nonumber \\
\vert\hat{\bf 1} \rangle_{L} & = & (\vert \tilde{0}0000 \rangle - \vert \tilde{1}1111 \rangle)_{12345} \otimes \vert \b{1}0001 \rangle_{1'2'3'4'5'} , \nonumber \\
\vert \hat{\bf 2} \rangle_{L} & = & (\vert \tilde{0}0010 \rangle + \vert \tilde{1}1101 \rangle)_{12345} \otimes \vert \b{1}0010 \rangle_{1'2'3'4'5'} , \nonumber \\
\vert \hat{\bf 3} \rangle_{L} & = & (\vert \tilde{0}0010 \rangle - \vert \tilde{1}1101 \rangle)_{12345} \otimes \vert \b{1}0011 \rangle_{1'2'3'4'5'} , \nonumber \\
\vdots & \vdots & \vdots \nonumber \\
\vert \hat{\bf 28}  \rangle_{L} & = & (\vert \tilde{1}1100 \rangle + \vert \tilde{0}0011 \rangle)_{12345} \otimes \vert \b{0}1100 \rangle_{1'2'3'4'5'} , \nonumber \\
\vert \hat{\bf 29}  \rangle_{L} & = & (\vert \tilde{1}1100 \rangle - \vert \tilde{0}0011 \rangle)_{12345} \otimes \vert \b{0}1101 \rangle_{1'2'3'4'5'} , \nonumber \\
\vert \hat{\bf 30} \rangle_{L} & = & (\vert \tilde{1}1110 \rangle + \vert \tilde{0}0001 \rangle)_{12345} \otimes \vert \b{0}1110 \rangle_{1'2'3'4'5'} , \nonumber \\
\vert \hat{\bf 31} \rangle_{L} & = & (\vert \tilde{1}1110 \rangle - \vert \tilde{0}0001 \rangle)_{12345} \otimes \vert \b{0}1111 \rangle_{1'2'3'4'5'} .  
\end{eqnarray}
}

\indent The next step is to apply a recovery operation to extract a state in the form of state (\ref {estadografo}). This operation can be performed by a unitary transformation in qubits $2, 3, 4, 5, 1', 2', 3', 4'$ and $5'$, which is described by 

{\footnotesize
\begin{equation}\label{oprec1}
\mathcal{U}_{r} = T_{1'5'4}Z_{5'4}T_{1'5'4}C_{2'2}C_{3'3'}C_{4'4'}C_{1'2}C_{1'3}C_{1'4}C_{1'5}
\end{equation}
}
where $ T_{1'5'4}$ is a Toffoli gate operation, and $Z_ {5'4}$ is a controlled Pauli $\sigma_{Z}$ operation, how descripted in Section \ref{erasurecode}.  Performing the calculations, one can easily verify that, after the operation $\mathcal{U}_{r}$, the system composed of ten qubits and the environment will be in state 

{\footnotesize
\begin{equation}\label{oprec2}
\left( \vert \tilde{0}0000 \rangle + \vert \tilde{1}1111 \rangle \right)_{12345} \otimes \vert \psi \rangle_{1'2'3'4'5'},
\end{equation}
}
where 
{\footnotesize
\begin{eqnarray}\label{estadografo2}
\vert \psi ' \rangle \triangleq \vert \psi \rangle_{1'2'3'4'5'} & = & \lambda_{0}  \vert \b{1}0000  \rangle  + \lambda_{1}  \vert \b{1}0001  \rangle  + \lambda_{2}  \vert \b{1}0010  \rangle + \lambda_{3}  \vert \b{1}0011  \rangle + \lambda_{4}  \vert \b{1}0100  \rangle  + \lambda_{5}  \vert \b{1}0101  \rangle \nonumber \\
& &   + \lambda_{6}  \vert \b{1}0110  \rangle  + \lambda_{7}  \vert \b{1}0111  \rangle + \lambda_{8}  \vert \b{1}1000  \rangle + \lambda_{9}  \vert \b{1}1001  \rangle  + \lambda_{10} \vert \b{1}1010  \rangle  + \lambda_{11}  \vert \b{1}1011  \rangle \nonumber \\
& &  + \lambda_{12} \vert \b{1}1100  \rangle +  \lambda_{13} \vert \b{1}1101  \rangle  + \lambda_{14} \vert \b{1}1110  \rangle + \lambda_{15} \vert \b{1}1111  \rangle + \lambda_{16} \vert \b{0}0000  \rangle   \nonumber \\
& &  + \lambda_{17} \vert \b{0}0001  \rangle + \lambda_{18} \vert \b{0}0010  \rangle + \lambda_{19} \vert \b{0}0011  \rangle + \lambda_{20} \vert \b{0}0100  \rangle + \lambda_{21} \vert \b{0}0101  \rangle  \nonumber \\
& & + \lambda_{22} \vert \b{0}0110  \rangle  + \lambda_{23} \vert \b{0}0111  \rangle + \lambda_{24} \vert \b{0}1000  \rangle  + \lambda_{25} \vert \b{0}1001  \rangle  + \lambda_{26} \vert \b{0}1010  \rangle \nonumber \\
& &  + \lambda_{27} \vert \b{0}1011  \rangle + \lambda_{28}  \vert \b{0}1100  \rangle + \lambda_{29} \vert \b{0}1101  \rangle + \lambda_{30} \vert \b{0}1110  \rangle  + \lambda_{31}  \vert \b{0}1111  \rangle .
\end{eqnarray} 
}

\indent From Eqs. (\ref{redund3}) and (\ref{oprec2}), one can note that the operators (\ref{operdec1}) and (\ref{oprec1}) perform a disentangling operation, which has made the five qubits $1', 2', 3', 4'$ and $5'$ no longer entangled with the remaining system, i.e., the five qubits 1, 2, 3, 4, 5 and the environment. Although the five qubits 1, 2, 3, 4, and 5 are entangled with the environment, the information, originally carried by qubits 1, 2, 3, 4 and 5, has been completely transferred into the five qubits $1', 2', 3', 4'$ and $5'$, and the state is reconstructed directly from the five qubits $1', 2', 3', 4'$ and $5'$. 

\indent The operations of decoding and recovery in case qubit 2 or 3 or 4 or 5 or $1'$ or $2'$ or $3'$ or $4'$ or $5'$  suffer erasure are similar to those used - see Table \ref{opercodGHZ} and Table \ref{operdecodGHZ}. 
{
\begin{table}[htb]
\centering
\caption{Decoding operators to protect against a quantum erasure.}
\label{opercodGHZ}
{\begin{tabular}{cc} \hline
Qubit positions & Decoding Operator \\ \hline
1, 2, 3, 4, 5 &  $\mathsf{ \mathcal{U}_{d} = H_{5'}C_{5'4'}C_{5'3'}C_{5'2'}C_{5'1'} }$  \\ 
$1', 2', 3', 4', 5'$ &  $\mathsf{\mathcal{U}_{d} = H_{5}C_{54}C_{53}C_{52}C_{51} }$  \\ \hline
\end{tabular}}
\end{table}
\indent \\
\begin{table}[htb]
\centering
\caption{Recovery operators to protect against a quantum erasure.}
\label{operdecodGHZ}
{\begin{tabular}{cc} \hline
Qubit positions  & Recovery Operator \\ \hline
1 &   $\mathsf{\mathcal{U}_{r} = T_{1'5'4}Z_{5'4}T_{1'5'4}C_{2'2}C_{3'3'}C_{4'4'}C_{1'2}C_{1'3}C_{1'4}C_{1'5} }$ \\ 
2 &  $\mathsf{\mathcal{U}_{r} = T_{2'5'3}Z_{5'3}T_{2'5'3}C_{1'1}C_{3'3'}C_{4'4'}C_{2'1}C_{2'3}C_{2'4}C_{2'5} }$ \\ 
3 &  $\mathsf{\mathcal{U}_{r} = T_{3'5'2}Z_{5'2}T_{3'5'2}C_{1'1}C_{2'2}C_{4'4}C_{3'1}C_{3'2}C_{3'4}C_{3'5} }$ \\ 
4 &  $\mathsf{\mathcal{U}_{r} = T_{4'5'1}Z_{5'1}T_{4'5'1}C_{1'1}C_{2'2}C_{3'3}C_{4'1}C_{4'2}C_{4'3}C_{4'5} }$ \\ 
5 &  $\mathsf{\mathcal{U}_{r} = Z_{5'4}C_{1'1}C_{2'2}C_{3'3}C_{4'4} }$ \\ 
$1'$ &  $\mathsf{\mathcal{U}_{r} = T_{154'}Z_{54'}T_{154'}C_{22'}C_{33'}C_{44'}C_{12'}C_{13'}C_{14'}C_{15'} }$ \\ 
$2'$ &  $\mathsf{\mathcal{U}_{r} = T_{253'}Z_{53'}T_{253'}C_{11'}C_{33'}C_{44'}C_{21'}C_{23'}C_{24'}C_{25'} }$ \\ 
$3'$ &  $\mathsf{\mathcal{U}_{r} = T_{352'}Z_{52'}T_{352'}C_{11'}C_{22'}C_{44'}C_{31'}C_{32'}C_{34'}C_{35'} }$ \\ 
$4'$ &  $\mathsf{\mathcal{U}_{r} = T_{451'}Z_{51'}T_{451'}C_{11'}C_{22'}C_{33'}C_{41'}C_{42'}C_{43'}C_{45'} }$ \\ 
$5'$ &  $\mathsf{\mathcal{U}_{r} = Z_{54'}C_{11'}C_{22'}C_{33'}C_{44'} }$ \\ \hline
\end{tabular}}
\end{table}
}
\indent Once you have retrieved the information which passed through the erasure channel, you must apply the decoder to the external code (quantum graph code) in order to correct the occurrence of a computational error to retrieve the information originally protected. 

\begin{figure}[h]
               \centering
               \includegraphics[scale=0.2]{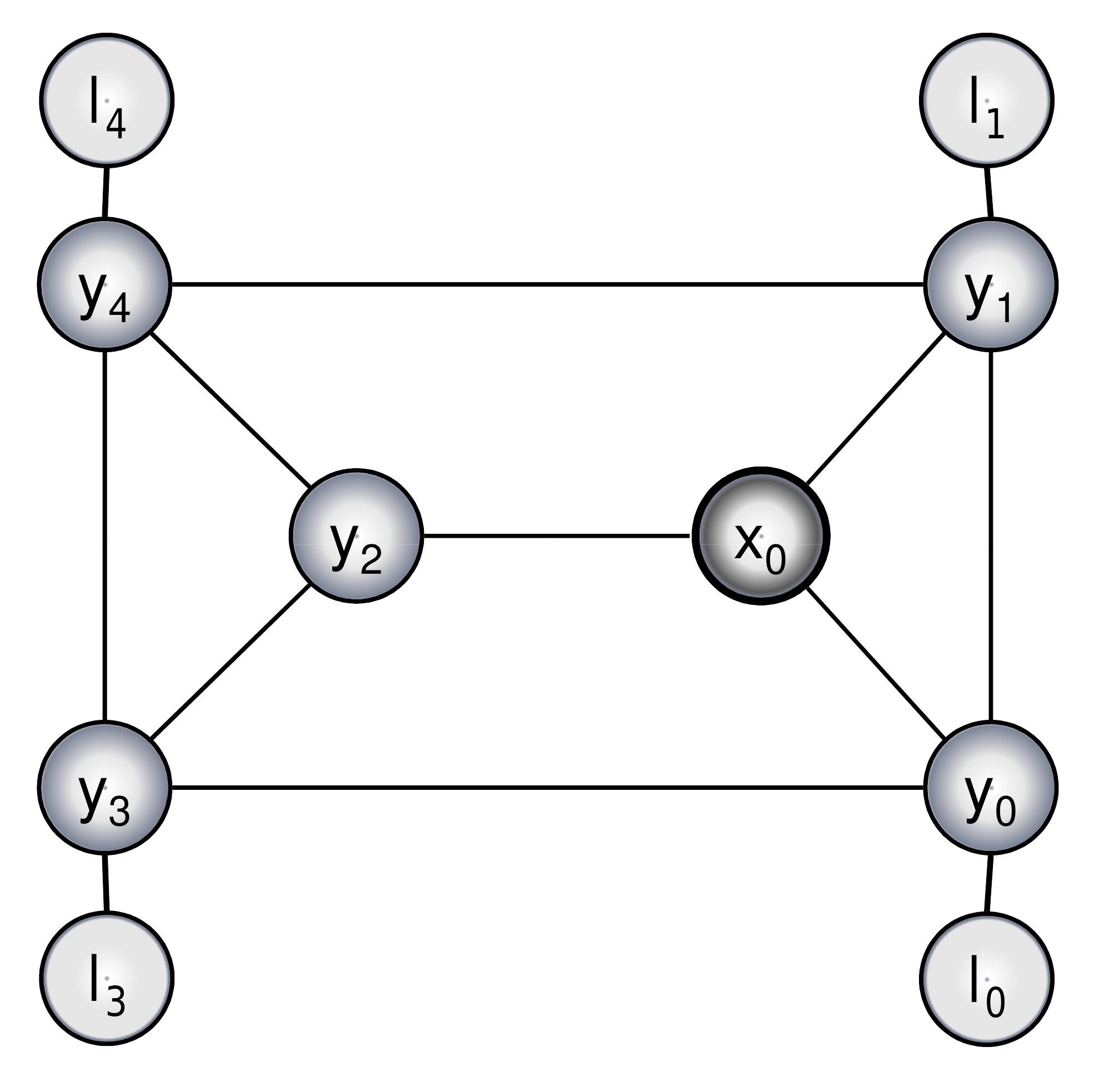}
               \caption{\footnotesize The 3-regular graph for the code [[5,1,3]] with syndrome vertices.}
               \label{figure:g3rsind}
  \end{figure}

\indent Thus, the graph used to decode the state $\vert \psi ' \rangle$, Eq. (\ref{estadografo2}), satisfying the Definition \ref{condgraphdecod} shall be as shown in Figure \ref{figure:g3rsind}.
  
\indent Take a graph that generates a encoding making use of Lemma \ref{defcodgraf} and by inserting vertices syndromes satisfies Definition \ref{condgraphdecod}. Thus, considering the graph shown in Figure \ref{figure:g3rsind}, which satisfies the Definition \ref{condgraphdecod}, the Eq. (\ref{iqftgrafo}) take the following form (the normalization factors are omitted): 
{\footnotesize
\begin{equation}\label{invg3r}
\mathcal{T}( \vert y_{0}y_{1}y_{2}y_{3}y_{4} \rangle ) = \sum_{l_{0}=0}^{1}\sum_{l_{1}=0}^{1}\sum_{l_{3}=0}^{1}\sum_{l_{4}=0}^{1} \sum_{x_{0}=0}^{1} e^{-(\pi i) \left( \theta \right)} \vert l_{0}l_{1}l_{3}l_{4}x_{0} \rangle,
\end{equation}
}
where 

{\footnotesize
$$
\theta=x_{0}y_{0}+x_{0}y_{1}+x_{0}y_{2}+y_{0}y_{1}+y_{0}y_{3}+y_{1}y_{4}+y_{2}y_{3}+y_{2}y_{4}+y_{3}y_{4}+y_{0}l_{0}+y_{1}l_{1}+y_{3}l_{3}+y_{4}l_{4}.
$$
}

\indent The inverse of state $\vert \psi ' \rangle$ - Eq. (\ref{estadografo2}) - will be given as 
{\footnotesize
\begin{eqnarray}\label{coderrografo}
\mathcal{R}(\vert \psi ' \rangle) & = & \lambda_{0}  \mathcal{T}(\vert \b{1}0000  \rangle)  + \lambda_{1}  \mathcal{T}(\vert \b{1}0001  \rangle)  + \lambda_{2}  \mathcal{T}(\vert \b{1}0010  \rangle) + \lambda_{3}  \mathcal{T}(\vert \b{1}0011  \rangle)  \nonumber \\
& &  + \lambda_{4}  \mathcal{T}(\vert \b{1}0100  \rangle)  + \lambda_{5} \mathcal{T}(\vert \b{1}0101  \rangle) + \lambda_{6}  \mathcal{T}(\vert \b{1}0110  \rangle)  + \lambda_{7}  \mathcal{T}(\vert \b{1}0111  \rangle)  \nonumber \\
& & + \lambda_{8} \mathcal{T}(\vert \b{1}1000  \rangle) + \lambda_{9}  \mathcal{T}(\vert \b{1}1001  \rangle)  + \lambda_{10} \mathcal{T}(\vert \b{1}1010  \rangle)  + \lambda_{11} \mathcal{T}(\vert \b{1}1011  \rangle) \nonumber \\
& &  + \lambda_{12}\mathcal{T}(\vert \b{1}1100  \rangle) +  \lambda_{13}\mathcal{T}(\vert \b{1}1101  \rangle)  + \lambda_{14} \mathcal{T}(\vert \b{1}1110  \rangle)  + \lambda_{15} \mathcal{T}(\vert \b{1}1111  \rangle) \nonumber \\
& &  + \lambda_{16} \mathcal{T}(\vert \b{0}0000  \rangle)  + \lambda_{17} \mathcal{T}(\vert \b{0}0001  \rangle) + \lambda_{18} \mathcal{T}(\vert \b{0}0010  \rangle) + \lambda_{19} \mathcal{T}(\vert \b{0}0011  \rangle) \nonumber \\
& & + \lambda_{20} \mathcal{T}(\vert \b{0}0100  \rangle) + \lambda_{21} \mathcal{T}(\vert \b{0}0101  \rangle) + \lambda_{22} \mathcal{T}(\vert \b{0}0110  \rangle)  + \lambda_{23} \mathcal{T}(\vert \b{0}0111  \rangle) \nonumber \\
& & + \lambda_{24} \mathcal{T}(\vert \b{0}1000  \rangle)  + \lambda_{25} \mathcal{T}(\vert \b{0}1001  \rangle) + \lambda_{26} \mathcal{T}(\vert \b{0}1010  \rangle) + \lambda_{27}\mathcal{T}(\vert \b{0}1011  \rangle)  \nonumber \\
& &  + \lambda_{28}  \mathcal{T}(\vert \b{0}1100  \rangle)  + \lambda_{29} \mathcal{T}(\vert \b{0}1101  \rangle) + \lambda_{30} \mathcal{T}(\vert \b{0}1110  \rangle)  + \lambda_{31}  \mathcal{T}(\vert \b{0}1111  \rangle) . 
\end{eqnarray} 
}

\indent Thus, to perform the decoding operation above you must first obtain the inverse of each one of the states of the base making use of the Eq. (\ref {invg3r}). 

\indent Substituting in the Eq.  (\ref{coderrografo}) the results of the inverses, making use of the Eq. (\ref {invg3r}), and multiplying them by their respective $\lambda_{i}$, being $i = 0, \ldots, 31$, it is obtained as result 
{\footnotesize
\begin{eqnarray}\label{finaldecod}
\mathcal{R}(\vert \psi ' \rangle) & = &   \vert 01100  \rangle c(0) - \vert 01101 \rangle c(1) \nonumber \\
                                      & = &   \vert 0 \rangle \vert 1 \rangle \vert 1 \rangle \vert 0 \rangle ( c(0) \vert 0 \rangle - c(1) \vert 1 \rangle) . 
\end{eqnarray}
}

\indent Recalling that the first four qubits are of measurement (syndromes), one just has to watch them and to check which type of occurrence is shown in the table of syndromes (see Table 3), and therefore see which action must be performed in the fifth qubit to get the desired state. 
\begin{table}[htb] 
{\centering
\caption{Table of syndromes: quantum code via 3-regular graph}
\label{tabsind3r}
{\begin{tabular}{cccc} \hline
Syndrome bits $q_{1}q_{2}q_{3}q_{4}$  & Error (*) & State of $q_{5}$ & Correction operation (*) \\ \hline
0000      &     None   &   $c(0) \vert 0 \rangle + c(1) \vert 1 \rangle$         & None \\ 
0001      &        $S_{5}$    &   $c(0) \vert 0 \rangle + c(1) \vert 1 \rangle$      & None \\ 
0010      &        $S_{4}$    &   $c(0) \vert 0 \rangle + c(1) \vert 1 \rangle$      & None \\ 
0011      &        $B_{3}$    &   $c(0) \vert 0 \rangle - c(1) \vert 1 \rangle$       &  $S_{5}$  \\ 
0100      &        $S_{2}$    &   $c(0) \vert 0 \rangle + c(1) \vert 1 \rangle$      & None \\ 
0101      &        $B_{4}$    &   $c(0) \vert 1 \rangle + c(1) \vert 0 \rangle$      & $B_{5}$   \\ 
{\bf 0110} &      $\mathbf{B_{1}}$  & $\mathbf{c(0) \vert 0 \rangle - c(1) \vert 1 \rangle}$  & $\mathbf{S_{5}}$ \\ 
0111      &        $BS_{4}$    &   $- c(0) \vert 1 \rangle - c(1) \vert 0 \rangle$    & $SBS_{5}$ \\ 
1000      &        $S_{1}$    &   $c(0) \vert 0 \rangle + c(1) \vert 1 \rangle$      & None \\ 
1001      &        $B_{2}$    &   $c(0) \vert 0 \rangle - c(1) \vert 1 \rangle$        & $S_{5}$ \\ 
1010      &        $B_{5}$    &   $c(0) \vert 1 \rangle + c(1) \vert 0 \rangle$       & $B_{5}$ \\ 
1011      &        $BS_{5}$    &   $-c(0) \vert 1 \rangle - c(1) \vert 0 \rangle$    & $SBS_{5}$ \\ 
1100      &        $S_{3}$    &   $c(0) \vert 1 \rangle + c(1) \vert 0 \rangle$       & $B_{5}$ \\ 
1101      &        $BS_{2}$    &   $-c(0) \vert 0 \rangle + c(1) \vert 1 \rangle$    & $BSB_{5}$ \\ 
1110      &        $BS_{1}$    &   $-c(0) \vert 0 \rangle + c(1) \vert 1 \rangle$    &  $BSB_{5}$ \\ 
1111      &        $BS_{3}$    &   $-c(0) \vert 1 \rangle + c(1) \vert 0 \rangle$    & $BS_{5}$ \\ \hline
\end{tabular}}
}
{\scriptsize 
\begin{tabular}{clc}
\hspace{1cm } (*) &$B_{n}$ denotes bit-flip on nth qubit;&\\
\hspace{1cm } \ &$S_{n}$ denotes sign-flip or phase-flip on nth qubit;&\\
\hspace{1cm } \ &$BS_{n}$ denotes sign and bit flip on nth qubit;&\\
\hspace{1cm } \ &$BSB_{5}$ denotes bit-flip, sign-flip end bit-flip on nth qubit, respectively;&\\
\hspace{1cm } \ &$SBS_{5}$ denotes sign-flip, bit-flip and sign-flip on nth qubit, respectively.& 
\end{tabular}}
\end{table}

\indent Therefore, after observing Table \ref{tabsind3r} it will be seen that the error made is bit-flip and the operation to be performed is sign-flip (or phase-flip) on the fifth qubit (highlighted in bold in the table). 

\indent Likewise, if the state $\vert \psi ' \rangle$ suffer any alteration, displayed in the Error column of Table 3, the $\mathcal{R}$ will enable the decoding operation and recovered the desired state making use of the Eq. (\ref{invg3r}). 

\indent That way, the information is recovered properly to the end of this scheme of concatenation, protecting it against the occurrence of an erasure and a computational error.

\section{Conclusion and Discussion}\label{conclus}
We have shown the realization of a scheme of concatenation that allows to protect a qubit of information against both the occurrence of one erasure and computational errors. 

\indent Furthermore, we showed up in an explicit way how the realization of the process of encoding and decoding for quantum graph codes is carried out, and for this we had to develop an IQFT to the appropriate decoding scheme for this type of code. To illustrate, is presented in Table 3 all the syndromes to the decoding of the encoded state via the 3-regular graph used.

\indent We have also presented a scheme to protect against the occurrence of one erasure involving GHZ states, generalizing the operators proposed by Yang et al.\cite{8} For ilustrate this way of proctet against the occurrence of one erasure, we build operators of five-qubits for the stages of encoding and decoding (see Tables 1 and 2), making use the operators presented in the Section \ref{erasurecode}. 

\indent As the next step, we intend to build a circuit involving the entire concatenated code in order to examine the possibility of improvement in its realization.

\indent Also, it is interesting to analyse which classes of graphs are appropriate for this type of concatenation and analyze how operators for encoding, decoding and reco\-ve\-ry can be constructed to handle more than one erasure.

\section*{Acknowledgments}

The authors acknowledge the general support of the Institute for Studies on Quantum Computation and Quantum Information  (IQuanta) of the Federal University of Campina Grande. This work was partially supported by CNPq (309431/2006.9).


\end{document}